\documentclass[usenatbib]{mnras}
\usepackage{mathptmx,graphics,graphicx,amssymb,amsmath}
\usepackage{color,multicol,fixltx2e,natbib,epsfig}

\include{defn}

\def\sgras{Sgr A$^\star$}

\def\xmm{{\sl XMM-Newton}}
\def\nustar{{\sl NuSTAR}}

\def\swift{{\sl Swift}}
\def\chandra{{\sl Chandra}}

\begin{document}

\title{A Systematic Chandra study of Sgr A$^{\star}$: II. X-ray flare 
statistics}

\author[]{Qiang Yuan$^{1,2}$\thanks{E-mail:yuanq@pmo.ac.cn},
Q. Daniel Wang$^3$\thanks{E-mail:wqd@astro.umass.edu},
Siming Liu$^{1,2}$, Kinwah Wu$^4$\\
$^1$Key Laboratory of Dark Matter and Space Astronomy, Purple Mountain 
Observatory, Chinese Academy of Sciences, Nanjing 210008, China\\
$^2$School of Astronomy and Space Science, University of Science and 
Technology of China, Hefei 230026, China\\
$^3$Department of Astronomy, University of Massachusetts, 710 North
Pleasant St., Amherst, MA, 01003, USA\\
$^4$Mullard Space Science Laboratory, University College London, 
Holmbury St. Mary, Dorking, Surrey, RH5 6NT, UK}

\maketitle

\label{firstpage}

\begin{abstract}
The routinely flaring events from \sgras\ trace dynamic, high-energy 
processes in the immediate vicinity of the supermassive black hole. 
We statistically study temporal and spectral properties, as well as 
fluence and duration distributions, of the flares detected by the \chandra\ 
X-ray Observatory from 1999 to 2012. The detection incompleteness and 
bias are carefully accounted for in determining these distributions. 
We find that the fluence distribution can be well characterized by a 
power-law with a slope of $1.73^{+0.20}_{-0.19}$, while the durations
($\tau$ in seconds) by a log-normal function with a mean 
$\log(\tau)=3.39^{+0.27}_{-0.24}$ and an intrinsic dispersion 
$\sigma=0.28^{+0.08}_{-0.06}$. No significant correlation between the 
fluence and duration is detected. The apparent positive correlation, 
as reported previously, is mainly due to the detection bias (i.e., weak 
flares can be detected only when their durations are short). These 
results indicate that the simple self-organized criticality model has 
difficulties in explaining these flares. We further find that bright 
flares usually have asymmetric lightcurves with no statistically evident 
difference/preference between the rising and decaying phases in terms of
their spectral/timing properties. Our spectral analysis shows that although
a power-law model with a photon index of $2.0\pm0.4$ gives a satisfactory 
fit to the joint spectra of strong and weak flares, there is weak
evidence for a softer spectrum of weaker flares. This work demonstrates 
the potential to use statistical properties of X-ray flares to probe 
their trigger and emission mechanisms, as well as the radiation 
propagation around the black hole.
\end{abstract}


\section{Introduction}


Low-luminosity supermassive black holes (LL-SMBHs) represent the silent
majority ($\sim90\%$) of SMBHs in our Universe.
\sgras\ is in a rather steady low-luminosity state, referred to as the 
``quiescent state'', with peak emission in the sub-millimeter band. 
Occasionally there are substantial variations in the emission, known 
as flares, which are most prominent in the (near) infrared (NIR/IR) and 
X-ray bands \citep{2003Natur.425..934G,2001Natur.413...45B}. The spatial, 
spectral, and temporal decompositions of the X-ray emission of \sgras\ 
show that 1) the quiescent emission is mostly extended and the flaring
emission is point-like \citep{2003ApJ...591..891B,2013Sci...341..981W};
2) there is an additional point-like, super-soft quiescent component 
which are not accounted for by detected flares
\citep{2017MNRAS.466.1477R}; 3) the spectrum is optically thin thermal 
for the quiescent extended emission while featureless power-laws for flares 
\citep{2001Natur.413...45B,2012ApJ...759...95N,2013Sci...341..981W};
4) the rate of X-ray flares is about $1\sim2$ per day 
\citep{2015MNRAS.454.1525P,2016MNRAS.456.1438Y} or about $3$ per day 
after correcting for the detection threshold \citep{2017arXiv170408102M}, 
which is a factor of a few smaller than that of NIR/IR ones 
\citep{2006A&A...450..535E}.


The quiescent emission of \sgras\ can be explained in terms 
of the radiatively inefficient inflow/outflow model  
\citep{2003ApJ...598..301Y,2012MNRAS.426.3241N,2013Sci...341..981W,
2015ApJ...804..101Y,2017MNRAS.466.1477R}. The origin of the flares is,
however, still unclear. 
From the temporal spectral properties, crucial information regarding the 
radiative mechanisms associated with the flares can be extracted. However, 
existing studies tended to focus on individual strong flares detected with 
reasonably good counting statistics, mostly via observations made with 
\xmm\ \citep{2003A&A...407L..17P,2005ApJ...635.1095B,2006ApJ...644..198Y,
2008A&A...488..549P} and a few with \chandra\ \citep{2001Natur.413...45B,
2012ApJ...759...95N} and \nustar\ \citep{2014ApJ...786...46B,
2017MNRAS.468.2447P}. Only a few works studied the flare population, 
with limited flare samples \citep{2013ApJ...774...42N,
2017arXiv170508002Z}. Moreover, the spectral shape of such flares is 
often modeled by an absorbed power-law. Comparison among the photon indices 
($\Gamma$) obtained for various flares is therefore not straightforward, 
when $\Gamma$ is strongly correlated with the foreground absorption column 
density $N_{\rm H}$ in the spectral fits. There could be differences in 
the modeling of such details due to adoption of different versions of 
the absorption cross-sections, dust absorption/scattering, and/or metal 
abundance pattern. For bright flares detected by \chandra, pile-up effects, 
which include the grade migration \citep{2001ApJ...562..575D}, can be 
problematic, as they cause distortion in the spectra data. Whether or not, 
and/or how the pile-up is treated can therefore affect the values of the 
photon indices when fitting the spectral data. With these in consideration, 
one finds that essentially all flares can be consistently characterized 
with a power law of $\Gamma\simeq2$ and $N_H \simeq 1.5 \times 10^{23} 
{\rm~cm^{-2}}$ of neutral material \citep{2008A&A...488..549P,
2012ApJ...759...95N}. This column density would be slightly smaller when 
dust scattering is accounted for separately. Nevertheless, the studies
of \nustar\ flares which extended the spectral coverage beyond 10~keV 
\citep[up to about 70 keV;][]{2014ApJ...786...46B} and \swift\ ones 
\citep{2013ApJ...769..155D}, do sometimes show that they may have 
different photon indices (e.g., $\Gamma \sim 3$). In this work, we extend 
the spectral analysis to relatively faint flares by both measuring
hardness ratios (HRs) of individual flares and fitting to stacked data. 

Flare statistics, on the other hand, may provide insights into the driving 
mechanism and how flares are triggered. It has been argued that flares are 
associated with the ejection of plasma blobs triggered by magnetic 
reconnection \citep[e.g.][]{2006ApJ...644..198Y}. One of the magnetic 
reconnection scenarios is that the system shows characteristics of 
self-organized criticality (SOC). In it, a critical state is reached 
gradually by nonlinear energy buildup, followed by an avalanche energy 
release, which manifests as a flaring event
\citep[e.g.,][]{1986JGR....9110412K,1987PhRvL..59..381B}. 
In such a SOC flaring model, if the system is scale-free, the total energy 
released in the flare, the peak rate of energy dissipation, and the flaring 
time duration should all obey a power-law distribution, and the slopes of 
these three power laws are determined by the effective geometric dimension 
of the system \citep{2012A&A...539A...2A,2016SSRv..198...47A}. SOC models 
have been applied to explain the statistics of flares in the Sun 
\citep[e.g.,][]{1991ApJ...380L..89L,2011SoPh..274...99A}, and in astrophysical 
black-hole systems \citep{2013NatPh...9..465W,2015ApJ...810...19L},   
The 3-Ms data of \sgras\ obtained in the \chandra\ X-ray Visionary Project 
(XVP) \citep{2013ApJ...774...42N} have shown that the X-ray flaring 
statistics of the source are consistent with those predicted by SOC models 
with a spatial dimension $S=3$ \citep{2015ApJS..216....8W,2015ApJ...810...19L}. 
However, the analyses might be limited by a relatively small sample of 
flares with narrow fluence range and by lacking a proper account for 
incompleteness and bias in the flare detection, the results obtained should 
be taken with caution.

\citet[][hereafter Paper I]{2016MNRAS.456.1438Y} have presented a  
systematical search for X-ray flares in 84 \chandra\ observations of 
\sgras. Fourty-six of these observations were taken before 2012, using 
the Advanced CCD Imaging Spectrometer - Imaging array (ACIS-I), while the 
other 38 in 2012, using the Advanced CCD Imaging Spectrometer - Spectroscopy 
array with the high energy transmission gratings (ACIS-S/HETG0, where ``0'' 
refers to the non-dispersed zeroth order). \chandra\ observations taken 
after 2012 are not included in the search because of the varying appearance 
of the X-ray bright magnetar, SGR J1745-2900 \citep{2013ApJ...770L..24K}, 
just $2.4''$ away from \sgras, which complicates the detection and 
statistical analysis of \sgras\ flares. With an improved unbinned likelihood 
method, the search finds a total of 82 flares in the $\sim4.5$ Ms 
observations, about 1/3 of which are newly detected ones (see 
Tables~\ref{table:flare-I} and ~\ref{table:flare-S} for a sub-sample with
relatively low pile-up effect). These two \chandra\ samples of \sgras\ 
flares form the base for the statistical analysis presented here. 
In addition, the detection incompleteness, uncertainty and bias are 
carefully studied for the first time, which is especially important for 
a statistical analysis including weak flares close to the detection 
threshold, as is the case for the work reported here. We adopt the 
detection response matrices, as obtained in Paper I, to better 
characterize the detection effects on the flare statistics.

\begin{table*}
\centering
\caption{Properties of the ACIS-I flares used for spectral analysis.}
\begin{tabular}{ccrccc}
\hline \hline
FlareID & $\log(F/{\rm cts})$ & $\log(\tau/{\rm ks})$ & $t_{\rm start}$ & $t_{\rm end}$ & $F_{\rm pileup}$ \\
    &  &  & (ks) & (ks)  &  \\
\hline
I1  & $1.07\pm0.23$ &  $0.45\pm0.20$ & 54270.053 & 54274.283 & 1.00 \\
I2  & $1.35\pm0.12$ &  $0.03\pm0.12$ & 89000.851 & 89002.455 & 0.93 \\
I3  & $1.00\pm0.26$ &  $0.14\pm0.29$ & 130520.43 & 130522.51 & 1.00 \\
I4  & $0.82\pm0.30$ & $-0.04\pm0.41$ & 133277.53 & 133278.89 & 1.00 \\
I5  & $1.85\pm0.10$ &  $0.80\pm0.09$ & 138651.24 & 138659.31 & 1.00 \\
I6  & $1.72\pm0.09$ &  $0.60\pm0.11$ & 138771.38 & 138777.35 & 0.98 \\
I7  & $1.02\pm0.19$ &  $0.01\pm0.26$ & 138781.96 & 138783.49 & 1.00 \\
I8  & $1.49\pm0.12$ &  $0.64\pm0.14$ & 138805.22 & 138811.78 & 1.00 \\
I9  & $1.39\pm0.12$ &  $0.18\pm0.09$ & 138864.21 & 138866.47 & 0.95 \\
I10 & $1.09\pm0.18$ &  $0.23\pm0.21$ & 138877.64 & 138880.18 & 1.00 \\
I11 & $2.18\pm0.07$ &  $0.76\pm0.10$ & 139036.87 & 139044.73 & 0.92 \\
I12 & $0.94\pm0.24$ & $-0.14\pm0.61$ & 139464.54 & 139465.62 & 1.00 \\
I13 & $1.17\pm0.22$ &  $0.47\pm0.24$ & 172451.56 & 172455.98 & 1.00 \\
I14 & $0.92\pm0.21$ & $-0.20\pm0.28$ & 205542.87 & 205543.81 & 0.99 \\
I15 & $1.77\pm0.08$ &  $0.83\pm0.09$ & 239074.25 & 239084.39 & 1.00 \\
I16 & $1.15\pm0.17$ &  $0.30\pm0.21$ & 265566.39 & 265569.39 & 1.00 \\
I17 & $1.07\pm0.19$ &  $0.32\pm0.16$ & 275579.64 & 275582.78 & 1.00 \\
I18 & $1.14\pm0.15$ &  $0.30\pm0.16$ & 305152.20 & 305155.20 & 1.00 \\
I19 & $0.99\pm0.26$ & $-0.49\pm0.51$ & 326370.81 & 326371.29 & 0.91 \\
I20 & $1.14\pm0.21$ &  $0.51\pm0.26$ & 333497.06 & 333501.92 & 1.00 \\
I21 & $1.21\pm0.18$ &  $0.36\pm0.26$ & 333503.03 & 333506.47 & 1.00 \\
I22 & $1.78\pm0.12$ &  $0.55\pm0.19$ & 359001.19 & 359005.59 & 0.95 \\
I23 & $1.90\pm0.12$ &  $0.61\pm0.09$ & 359026.86 & 359032.09 & 0.94 \\
I24 & $1.28\pm0.14$ &  $0.39\pm0.13$ & 417781.80 & 417785.48 & 1.00 \\
\hline \hline
\end{tabular}\\
Note: Columns from left to right are: flare ID, logarithmic flare fluence,
logarithmic flare duration, start and end times from UT 1998-01-01 00:00:00,
which define the flare intervals, and pile-up correction factor.
\label{table:flare-I}
\end{table*}

\begin{table*}
\centering
\caption{Properties of the ACIS-S/HETG0 flares used for spectral analysis.}
\begin{tabular}{ccrccc}
\hline \hline
FlareID & $\log(F/{\rm cts})$ & $\log(\tau/{\rm ks})$ & $t_{\rm start}$ & $t_{\rm end}$ & $F_{\rm pileup}$ \\
    &  &  & (ks) & (ks)  &  \\
\hline
S1 & $1.94\pm0.05$ &  $0.49\pm0.05$ & 453264.94 & 453269.58 & 0.93 \\
S2 & $1.82\pm0.11$ &  $0.30\pm0.10$ & 453933.00 & 453935.77 & 0.92 \\
S3 & $1.85\pm0.06$ &  $0.59\pm0.05$ & 459317.44 & 459323.28 & 0.95 \\
S4 & $1.85\pm0.09$ &  $0.94\pm0.08$ & 459428.50 & 459438.92 & 1.00 \\
S5 & $2.09\pm0.04$ &  $0.60\pm0.03$ & 460110.73 & 460116.71 & 0.92 \\
S6 & $2.01\pm0.08$ &  $0.51\pm0.11$ & 460253.06 & 460257.06 & 0.92 \\
S7 & $2.24\pm0.06$ &  $0.82\pm0.10$ & 467370.02 & 467380.57 & 0.93 \\
\hline
S8 & $1.18\pm0.13$ &  $0.27\pm0.14$ & 445170.33 & 445173.11 & 1.00 \\
S9 & $1.37\pm0.11$ & $-0.10\pm0.09$ & 448630.66 & 448631.84 & 0.92 \\
S10& $1.38\pm0.11$ &  $0.12\pm0.10$ & 448633.82 & 448635.80 & 0.95 \\
S11& $1.37\pm0.11$ &  $0.19\pm0.10$ & 448638.60 & 448640.92 & 0.96 \\
S12& $1.55\pm0.08$ &  $0.59\pm0.08$ & 452260.13 & 452265.97 & 1.00 \\
S13& $1.43\pm0.11$ &  $0.40\pm0.14$ & 452746.05 & 452749.81 & 1.00 \\
S14& $1.36\pm0.11$ &  $0.67\pm0.13$ & 452774.14 & 452781.16 & 1.00 \\
S15& $1.41\pm0.18$ &  $0.86\pm0.58$ & 453136.68 & 453143.17 & 1.00 \\
S16& $1.27\pm0.12$ &  $0.47\pm0.13$ & 453168.52 & 453172.94 & 1.00 \\
S17& $1.10\pm0.21$ &  $0.64\pm0.32$ & 453192.47 & 453199.03 & 1.00 \\
S18& $1.00\pm0.18$ &  $0.32\pm0.19$ & 453821.66 & 453824.80 & 1.00 \\
S19& $1.11\pm0.15$ &  $0.20\pm0.15$ & 453937.72 & 453940.09 & 1.00 \\
S20& $1.06\pm0.17$ &  $0.47\pm0.15$ & 453944.22 & 453948.64 & 1.00 \\
S21& $1.56\pm0.08$ &  $0.74\pm0.07$ & 459039.34 & 459047.59 & 1.00 \\
S22& $1.16\pm0.14$ &  $0.43\pm0.12$ & 459057.69 & 459061.73 & 1.00 \\
S23& $1.39\pm0.11$ &  $0.09\pm0.14$ & 459176.29 & 459178.13 & 0.94 \\
S24& $1.03\pm0.15$ & $-0.09\pm0.14$ & 459217.17 & 459218.39 & 0.99 \\
S25& $1.36\pm0.11$ &  $0.02\pm0.10$ & 459380.52 & 459382.10 & 0.94 \\
S26& $1.47\pm0.09$ &  $0.04\pm0.08$ & 459508.28 & 459509.93 & 0.93 \\
S27& $0.95\pm0.18$ & $-0.29\pm0.19$ & 459605.82 & 459606.58 & 0.95 \\
S28& $1.27\pm0.13$ &  $0.55\pm0.14$ & 459860.71 & 459866.03 & 1.00 \\
S29& $0.96\pm0.28$ &  $0.55\pm0.33$ & 459873.61 & 459878.93 & 1.00 \\
S30& $1.41\pm0.10$ &  $0.73\pm0.09$ & 460040.91 & 460048.97 & 1.00 \\
S31& $0.86\pm0.20$ & $-0.05\pm0.23$ & 460268.82 & 460270.16 & 0.99 \\
S32& $1.60\pm0.08$ &  $0.13\pm0.05$ & 460452.53 & 460454.55 & 0.92 \\
S33& $1.57\pm0.10$ &  $1.24\pm0.09$ & 460482.85 & 460508.95 & 1.00 \\
S34& $1.37\pm0.10$ &  $0.52\pm0.13$ & 460539.33 & 460544.29 & 1.00 \\
S35& $1.30\pm0.12$ &  $0.40\pm0.13$ & 460781.60 & 460785.36 & 1.00 \\
S36& $1.22\pm0.15$ &  $0.54\pm0.17$ & 465968.67 & 465973.87 & 1.00 \\
S37& $1.40\pm0.10$ &  $0.38\pm0.09$ & 466057.06 & 466060.66 & 1.00 \\
S38& $0.74\pm0.24$ & $-0.14\pm0.25$ & 466827.00 & 466828.08 & 1.00 \\
S39& $1.20\pm0.13$ &  $0.34\pm0.19$ & 466970.77 & 466974.05 & 1.00 \\
S40& $1.19\pm0.24$ &  $0.88\pm0.34$ & 467413.12 & 467424.50 & 1.00 \\
S41& $1.59\pm0.12$ &  $0.38\pm0.10$ & 467529.97 & 467533.23 & 0.96 \\
S42& $1.66\pm0.08$ &  $0.84\pm0.10$ & 467965.49 & 467975.87 & 1.00 \\
S43& $0.83\pm0.19$ & $-0.03\pm0.19$ & 468004.79 & 468006.19 & 1.00 \\
S44& $1.49\pm0.10$ &  $0.51\pm0.14$ & 468076.64 & 468081.50 & 1.00 \\
\hline \hline
\end{tabular}\\
Note: Same as Table \ref{table:flare-I}. The central horizontal line 
separates the strong flares from the weak ones.
\label{table:flare-S}
\end{table*}

To provide further constraints on the nature of the flares, we 
statistically characterize their time profiles and spectral variations. 
There have been a few studies on such properties of a few individual 
bright flares \citep[e.g.][]{2001Natur.413...45B,2003A&A...407L..17P,
2005ApJ...635.1095B,2006ApJ...644..198Y,2008A&A...488..549P,
2012ApJ...759...95N,2013ApJ...769..155D,2014ApJ...786...46B,
2017MNRAS.468.2447P}. We extend these studies to relatively weak flares, 
e.g., via stacking analysis. 

The organization for the rest of this paper is as follows. In Section 2
we present the statistical analysis of the X-ray flares. The implications 
of our results in understanding the nature of the flares are briefly
discussed in Section 3. Finally we summarize our work in Section 4.

\section{Flare statistics}

\subsection{Fluence and duration distributions}

This analysis follows the approach of \citet{2015ApJ...810...19L} to 
characterize the probability distributions of the flare fluence ($F$) and 
duration ($\tau$). The distribution of $F$ is assumed to be a power-law, 
$P(F)=A\cdot F^{-\alpha}$, while $\tau$ follows a log-normal function,
$N(\log\tau;\mu,\sigma)$, in which $\mu=\log(B\cdot F^\beta)$ is the 
expected mean correlation with the fluence and $\sigma$ is the Gaussian 
width of $\log\tau$\footnote{This treatment is essentially the same
as adding an ``intrinsic'' error to the statistical one of $\log\tau$,
as done in Paper I.}. Hereafter we use $\log F$ and $\log\tau$ as 
variables. The joint {\sl intrinsic} probability distribution of the 
fluence ($\log F_i$) and duration ($\log\tau_i$) is then
\begin{eqnarray}
P(\log F_i,\log\tau_i)&=&P(\log F_i)\cdot P(\log\tau_i|\log F_i) \nonumber\\
&=&F_i\cdot\ln10\cdot P(F_i)\cdot N(\log\tau_i;\log B+\beta\log F_i,\sigma). 
\nonumber\\
\end{eqnarray}
The joint probability distribution of the {\sl detected}fluence ($\log F_d$) and 
duration ($\log\tau_d$)  is
\begin{equation}
P(\log F_d,\log\tau_d)=P(\log F_d,\log\tau_d;\,\log F_i,\log\tau_i)
\otimes P(\log F_i,\log\tau_i),\label{eq:prob2}
\end{equation}
where $\otimes$ means the convolution of $P(\log F_i,\log\tau_i)$ with 
$P(\log F_d,\log\tau_d;\,\log F_i,\log\tau_i)$, which is a redistribution 
matrix. It is obtained through Monte Carlo simulations for the two flare 
samples separately, accounting for the counting statistics and 
background-dependent detection incompleteness and bias (see Paper I). 
Individual flares are considered to be independent Poisson realizations. 
The logarithmic likelihood function of our $N_{d}$  detected flare is then
\citep{1979ApJ...228..939C}
\begin{equation}
\ln{\mathcal L}(\vec{\theta}|{\rm Data})=\sum_k^{N_{d}}\ln P(\log F^k_d,
\log\tau^k_d) - N_{\rm pred},\label{eq:like}
\end{equation}
where $\vec{\theta}=(A,\alpha,B,\beta,\sigma)$ represent the model 
parameters, the sum is over all the detections ($k=1,...,N_{d}$) and
\begin{equation}
N_{\rm pred}=\int\int P(\log F_d,\log\tau_d)\,{\rm d}\log F_d
\,{\rm d}\log\tau_d
\end{equation}
is the expected total number of flares. We use the Markov Chain Monte 
Carlo (MCMC) method to maximize Eq.~(\ref{eq:like}) and constrain the model 
parameters $\vec{\theta}$. Compared with Paper I, we improve the flare
statistical study through proper considerations of the Poisson fluctuation 
and the detection bias in a joint fit of the fluence distribution and the 
fluence-duration correlation.

\begin{figure*}
\centering
\includegraphics[width=1.05\columnwidth]{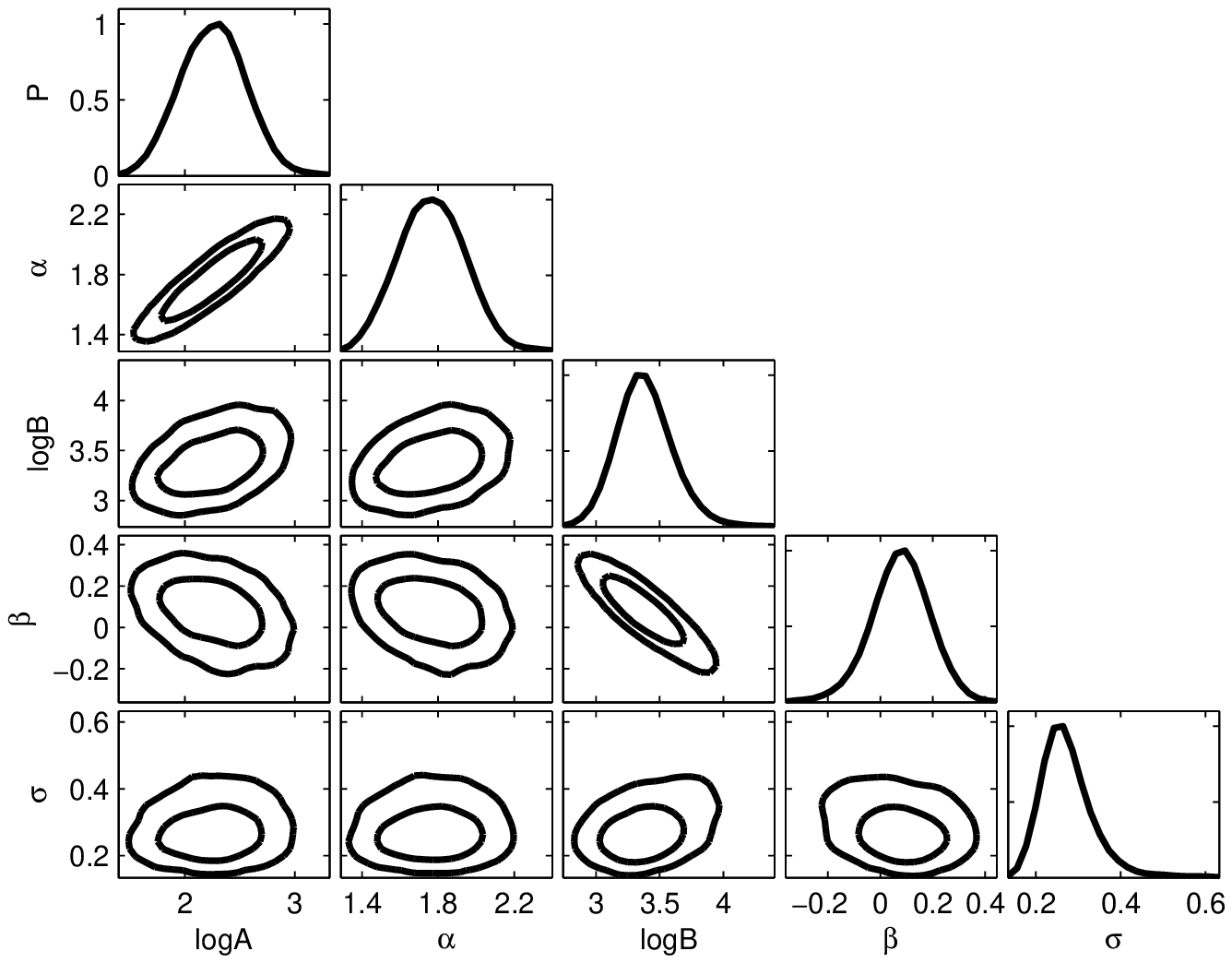}\hspace{-0.8cm}
\includegraphics[width=1.05\columnwidth]{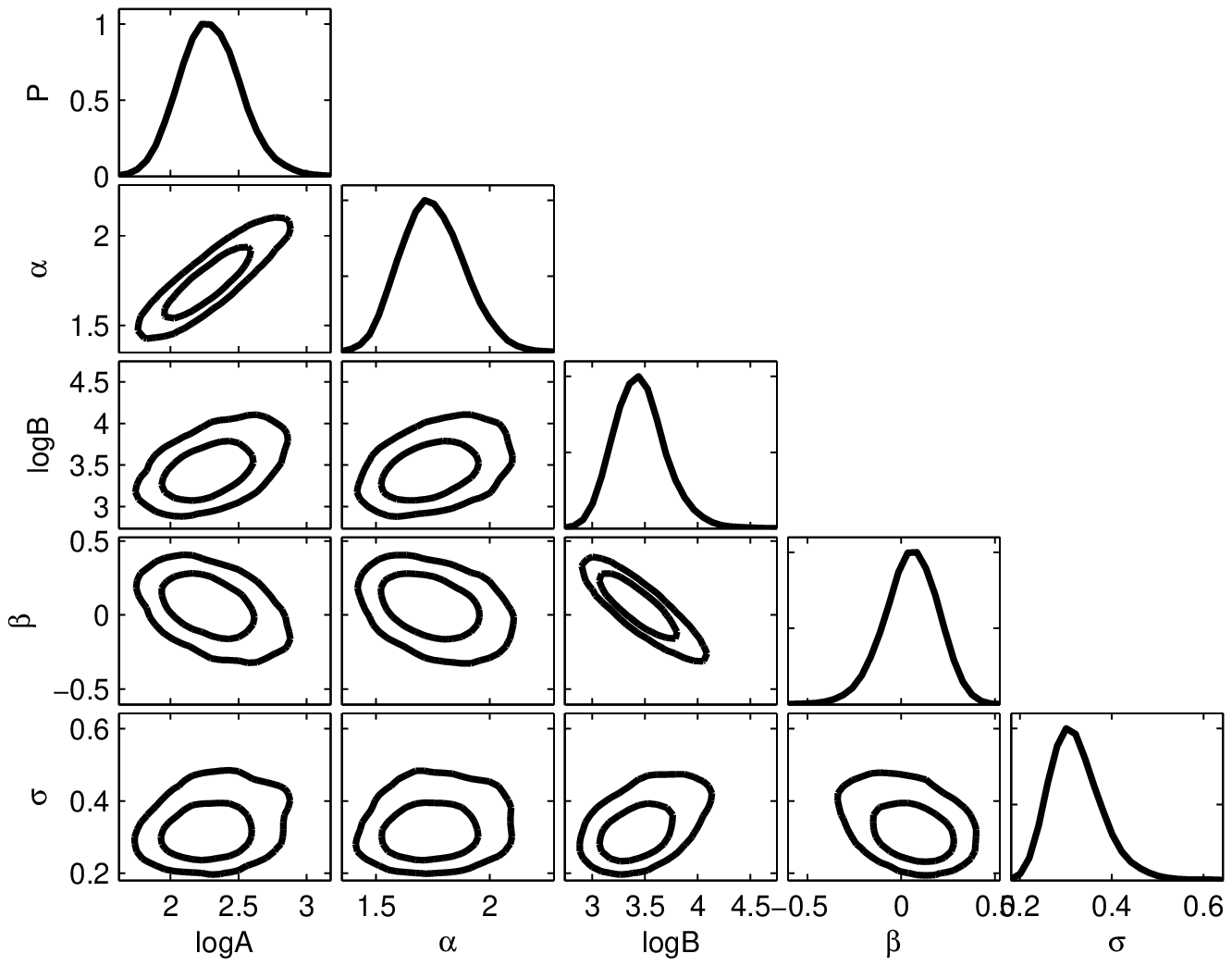}
\caption{Fitting 1-d (diagonal) probability distributions and 2-d 
(off-diagonal) contours at 68\% and 95\% confidence levels of the 
model parameters, $(\log A,\alpha,\log B,\beta,\sigma)$, for the 
ACIS-I (left) and -S/HETG0 (right) data.
}
\label{fig:para}
\end{figure*}

Table \ref{table:fit} gives the best-fit and posterior two-sided $95\%$ 
confidence ranges of the parameters. 
The corresponding 1-dimensional (1-d) and 2-dimensional (2-d) distributions 
of the fitting parameters are shown in Figure \ref{fig:para}. The parameters obtained for the ACIS-I and -S/HETG0 flares
are consistent with each other. 

\begin{table*}
\centering
\caption{The best-fit, posterior mean values and the 95\% limits of the 
logarithmic normalization ($\log A$) and power-law index ($\alpha$) of the 
fluence distribution, and the logarithmic normalization ($\log B$), power-law 
index ($\beta$), and dispersion width ($\sigma$) of the fluence-duration 
correlation (see \S~2.1).}
\begin{tabular}{ccccccccccccccc}
\hline \hline
    & \multicolumn{2}{c}{$\log A$} & & \multicolumn{2}{c}{$\alpha$} & & \multicolumn{2}{c}{$\log B$} & & \multicolumn{2}{c}{$\beta$} & & \multicolumn{2}{c}{$\sigma$} \\
    \cline{2-3} \cline{5-6} \cline{8-9} \cline{11-12} \cline{14-15}
    &  best & posterior mean  & & best & posterior mean  & & best & posterior mean  & & best & posterior mean  & & best & posterior mean \\
    &       & and 95\% limits & &      & and 95\% limits & &      & and 95\% limits & &      & and 95\% limits & &      & and 95\% limits \\
\hline
ACIS-I & 2.13 & $2.26_{-0.55}^{+0.56}$ & & 1.68 & $1.77_{-0.32}^{+0.33}$ & & 3.34 & $3.38_{-0.38}^{+0.46}$ & & 0.09 & $0.08_{-0.23}^{+0.21}$ & & 0.25 & $0.28_{-0.09}^{+0.15}$ \\
ACIS-S/HETG0 & 2.24 & $2.29_{-0.40}^{+0.46}$ & & 1.71 & $1.75_{-0.24}^{+0.28}$ & & 3.35 & $3.45_{-0.41}^{+0.53}$ & & 0.10 & $0.05_{-0.30}^{+0.26}$ & & 0.28 & $0.32_{-0.09}^{+0.13}$ \\
Joint fit & 2.22 & $2.23\pm0.29$ & & 1.72 & $1.73_{-0.19}^{+0.20}$ & & 3.38 & $3.39_{-0.24}^{+0.27}$ & & 0.09 & $0.08_{-0.17}^{+0.15}$ & & 0.28 & $0.28_{-0.06}^{+0.08}$ \\
\hline
\hline
\end{tabular}
\label{table:fit}
\end{table*}

\begin{figure*}
\centering
\includegraphics[width=\columnwidth]{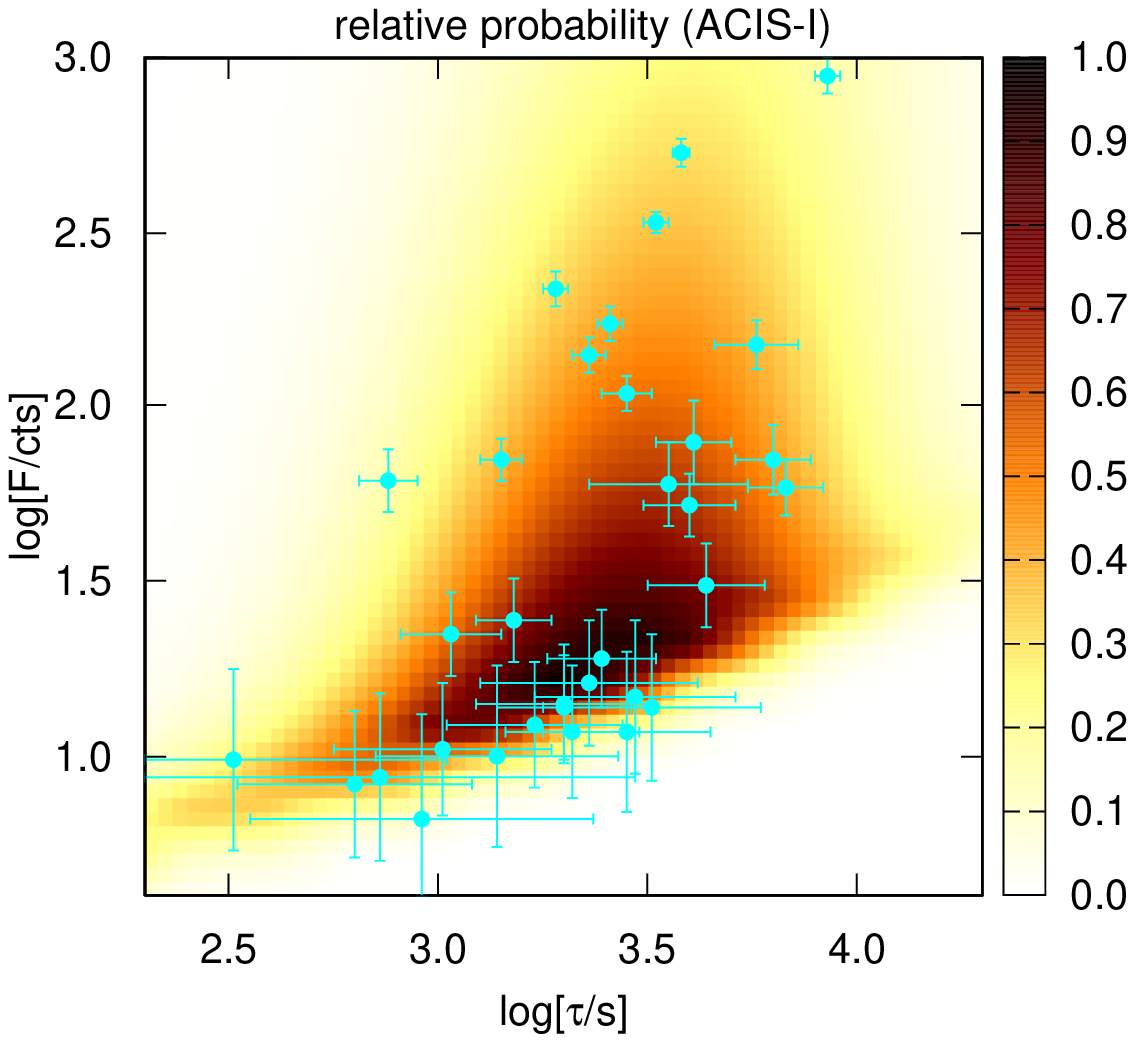}
\includegraphics[width=\columnwidth]{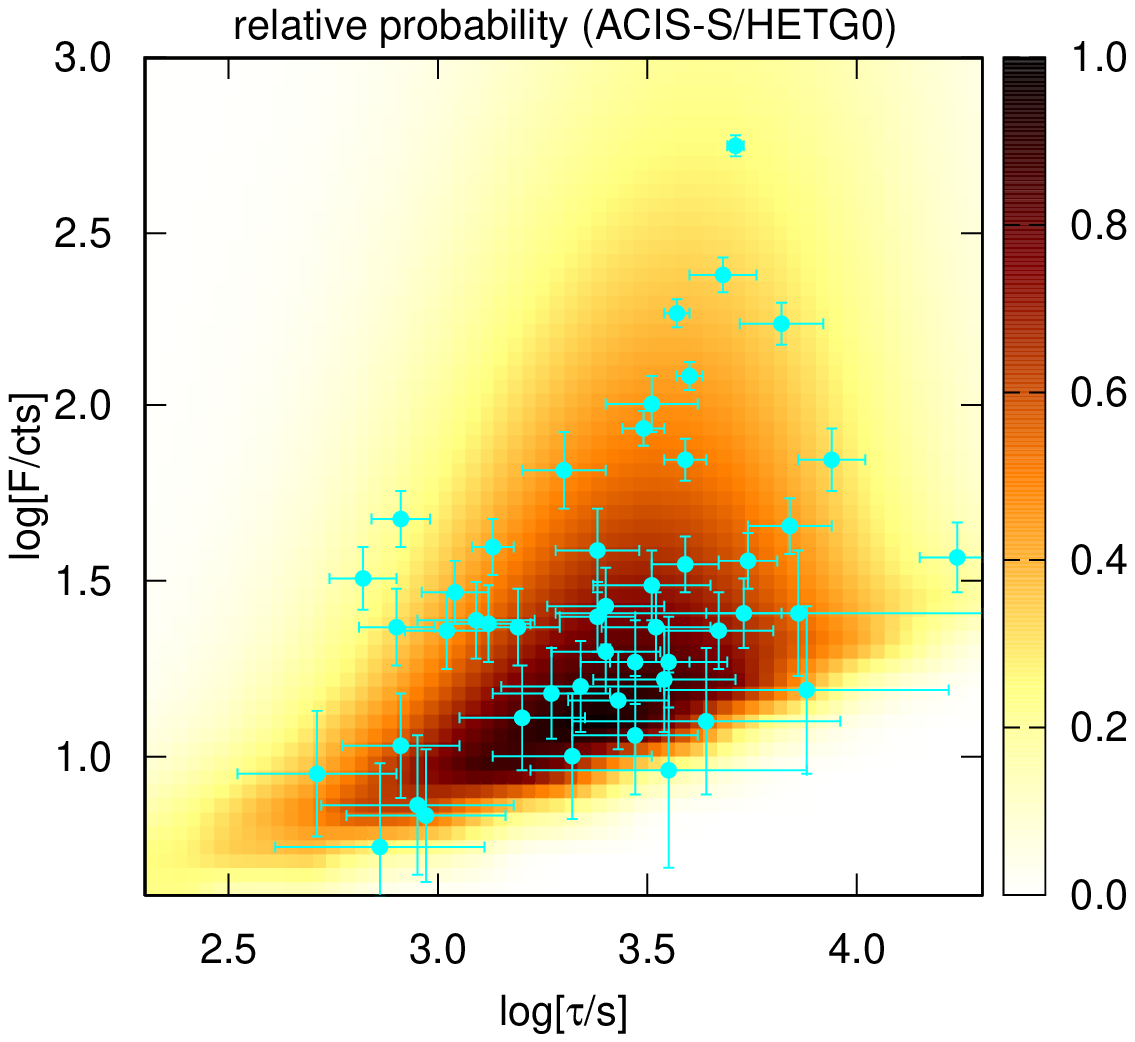}\\
\vspace*{-15mm}
\includegraphics[width=\columnwidth]{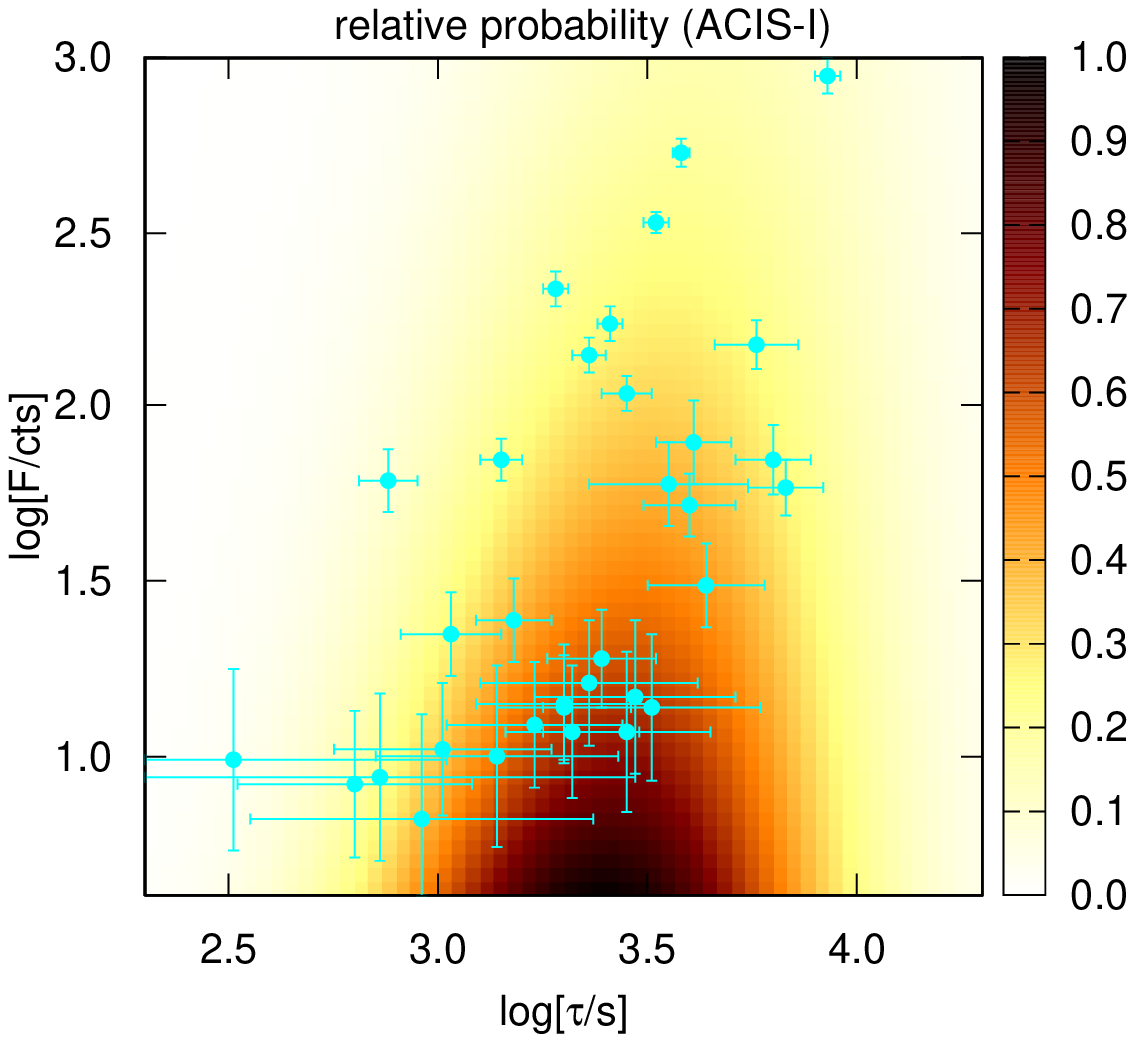}
\includegraphics[width=\columnwidth]{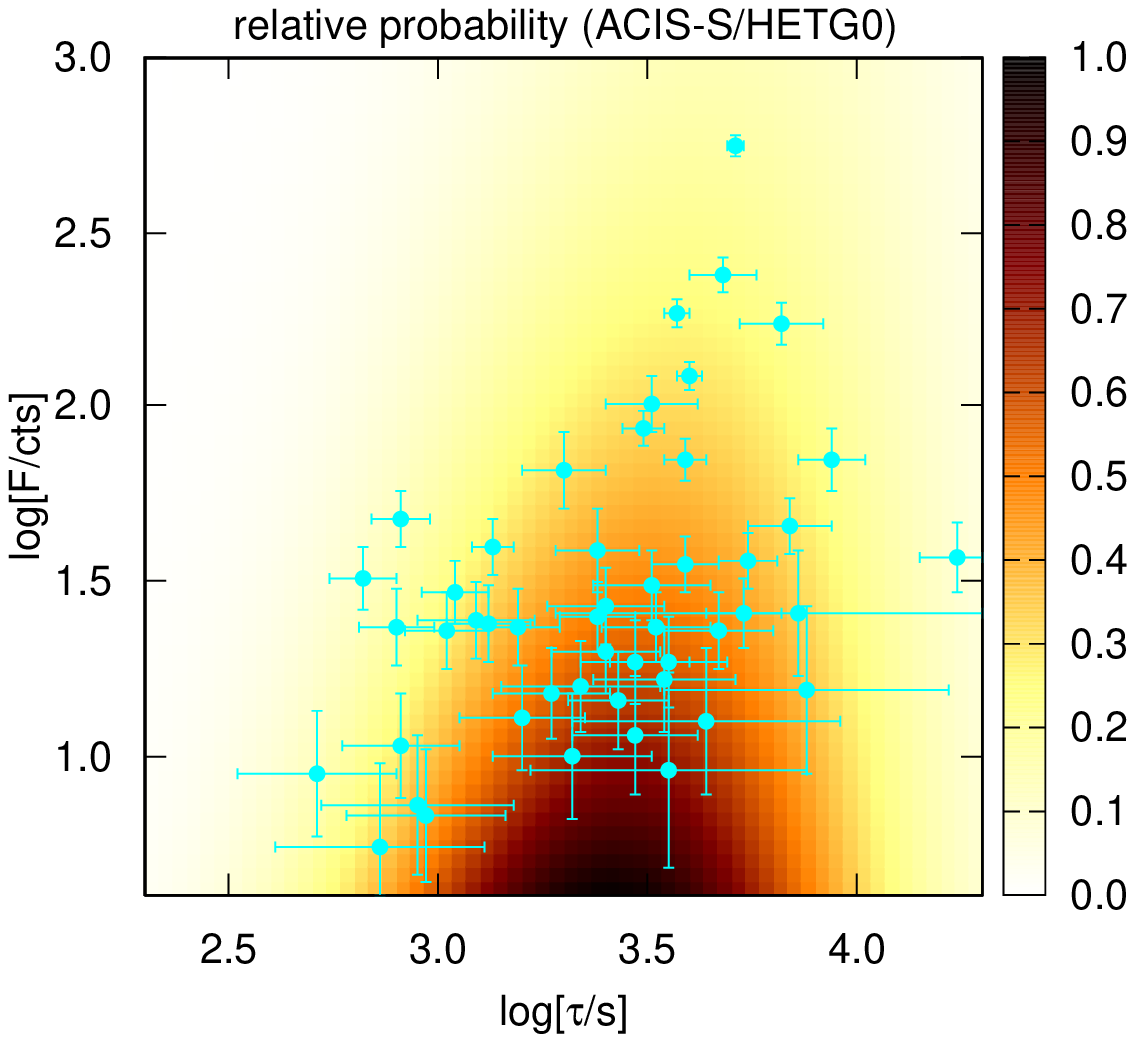}
\vspace*{-8mm}
\caption{The images in the top two panels show the relative probability 
distributions of the flare detection as a function of the fluence and 
duration, for the ACIS-I (left) and -S/HETG0 (right) samples. The 
overlaid data points are from our detected flares in the respective samples
(Paper I). For comparison, the images in the bottom two panels show the 
intrinsic probability distributions of flares without convolution with 
the redistribution matrices.
}
\label{fig:prob}
\end{figure*}

The top two panels of Figure \ref{fig:prob} show the detection probability 
distribution as a function of  $\log F_d$ and 
$\log\tau_d$ (Eq.~\ref{eq:prob2}) for the best-fit models of the two flare samples, respectively. As a comparison, we show in the bottom two panels the 
intrinsic probability distribution without the convolution with the 
detection redistribution matrix. It clearly shows how an apparent 
correlation can be obtained from an intrinsically nearly uncorrelated 
distribution between the fluence and duration. The detection redistribution 
matrix makes long duration, weak flares undetectable and the probability 
distribution wider.

\begin{figure*}
\centering
\includegraphics[width=1.0\columnwidth]{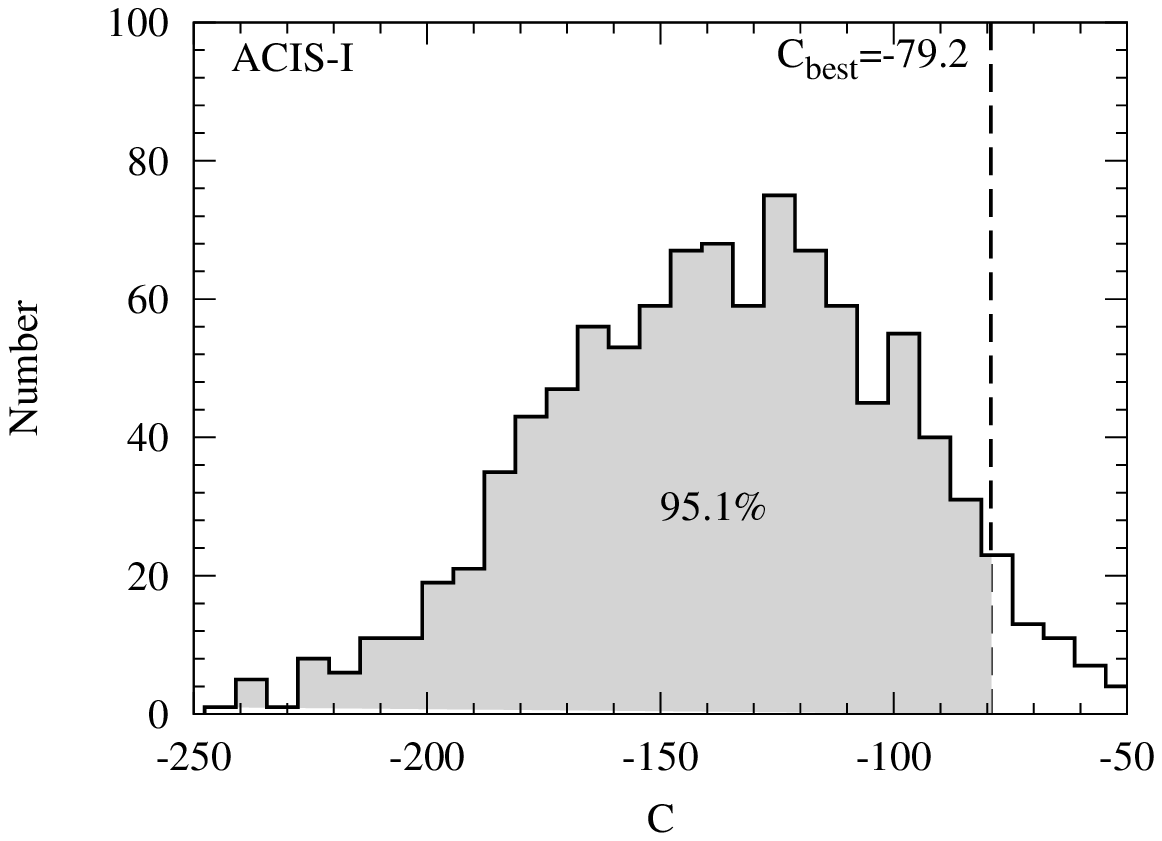}
\includegraphics[width=1.0\columnwidth]{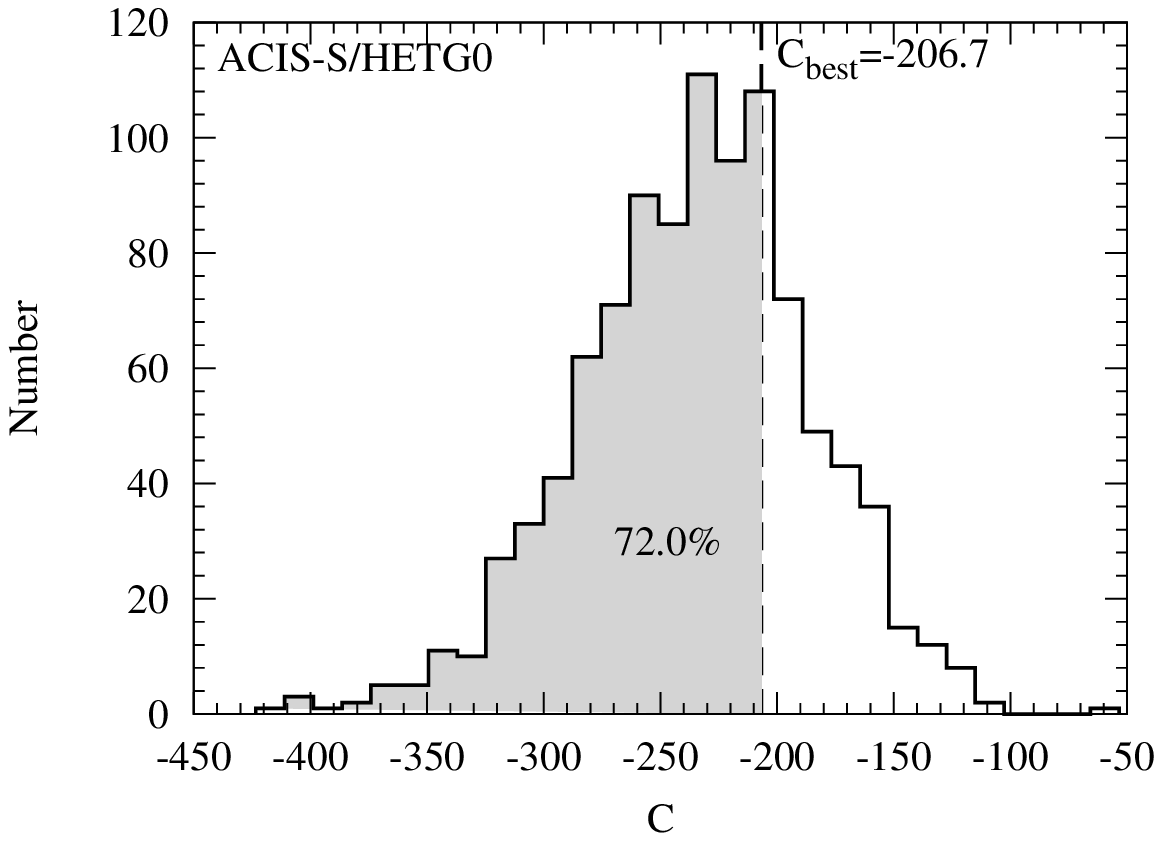}
\caption{Distributions of the $C$-statistic values (defined as
$-2\ln{\mathcal L}$) of the fits to the 1000 sets of statistically realized
flares, following the best-fit fluence-duration distributions, for the 
ACIS-I (left) and -S/HETG0 (right) samples. 
}
\label{fig:gof}
\end{figure*}

We assess the goodness of the fit to the detected flares from each of the 
two detected flare samples via bootstrapping sampling. Figure \ref{fig:gof} 
present the distributions of $C\equiv-2\ln{\mathcal L}$ from the fits to the 
1000 sets of bootstrapped flares, which are randomly realized from the 
best-fit model. The number fraction with $C$ smaller than that of the 
actual data ($C_{\rm best}=-206.7$) is $72\%$ for the ACIS-S/HETG0 
flares, suggesting that the data are well described by the model. 
The corresponding fraction is $95.1\%$ for the ACIS-I data, which means 
a slightly worse fitting.

We further jointly fit the two flare samples to improve the constraints on 
the model parameters. Since the effective area (exposure time) of the 
ACIS-I observations is on average a factor of $\sim2.6$ ($2.0$) larger 
(smaller) than that of the ACIS-S/HETG0 observations (Paper I), we expect 
to have $P(F_I)=P(F_S)\cdot{\rm d}F_S/{\rm d}F_I\cdot t_I/t_S$, and hence 
$A_I=A_S\cdot2.6^{\alpha-1}/2$, where the subscription ``$_I$'' (``$_S$'') 
stands for the ACIS-I (-S/HETG0) flares. Similarly for the fluence-duration 
correlation we have $B_I=B_S/2.6^{\beta}$. The joint fit significantly 
tightens the constraints on the model parameters, which are included in 
Table \ref{table:fit}. 

The power-law index of the fluence distribution, $\alpha\sim1.7$, 
is consistent with those found in \citet{2013ApJ...774...42N} and 
\citet{2015ApJ...810...19L}. But we find little intrinsic correlation 
between the fluence and duration ($\beta\sim0$), although an apparent 
correlation is present for the detected flares \citep[e.g., Figure 
\ref{fig:para}; Paper I;][]{2013ApJ...774...42N,2015ApJ...810...19L}. 
Such correlations are largely due to the detection bias and uncertainty, 
which were not fully accounted for previously.

\subsection{Flare time profiles}

We characterize the asymmetry properties of flare time profiles. In Paper 
I we used only the standard symmetric Gaussian profiles to approximate the 
flare lightcurves. Here we relax this approximation for those ``strong'' 
flares, each with fluence $F>50$ counts. We adopt a modified Gaussian 
function of varying width
\citep{2008ViS....47...66S}
\begin{equation}
\sigma(t)=\frac{2\sigma_0}{1+\exp[-\xi(t-t_0)]}.
\end{equation}
This function recovers to the standard Gaussian function with a constant 
width $\sigma_0$ when $\xi=0$. When $\xi>0$, the profile will be broader 
for $t>t_0$ and narrower for $t<t_0$, and vise versa when $\xi<0$. 

\begin{figure*}
\centering
\includegraphics[width=\columnwidth]{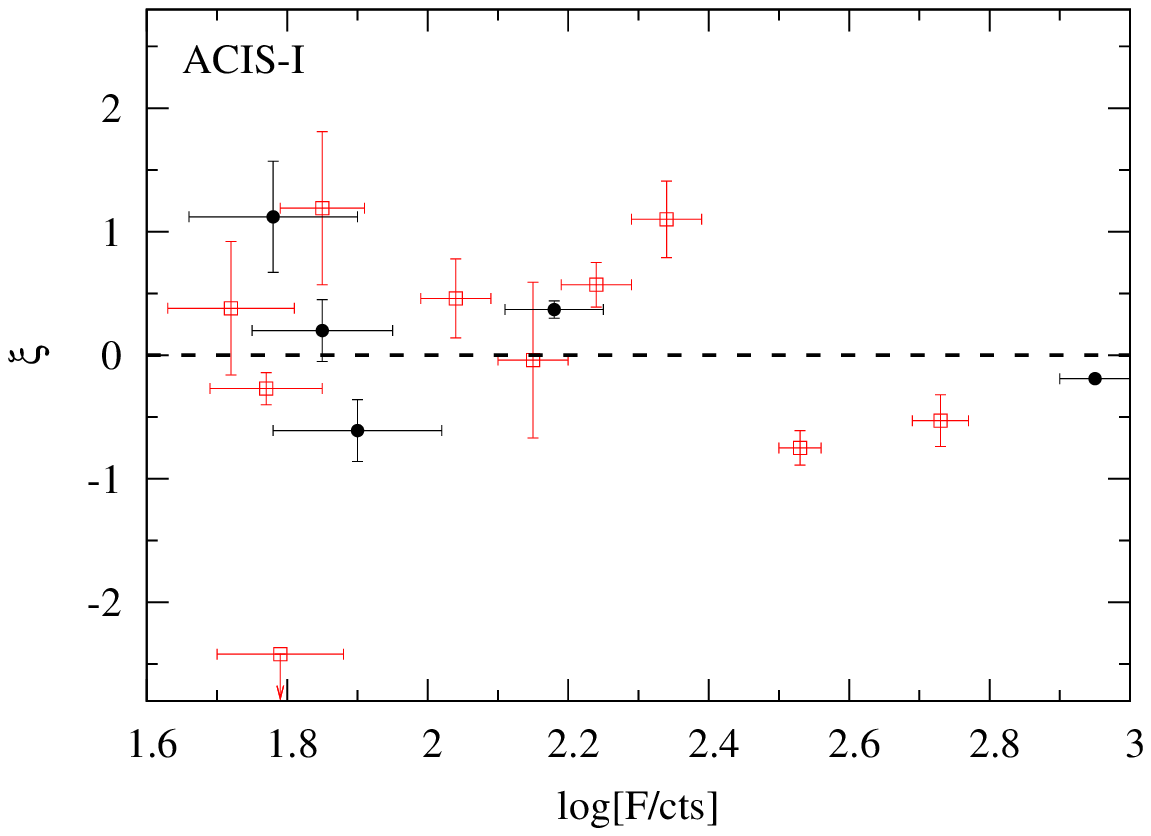}
\includegraphics[width=\columnwidth]{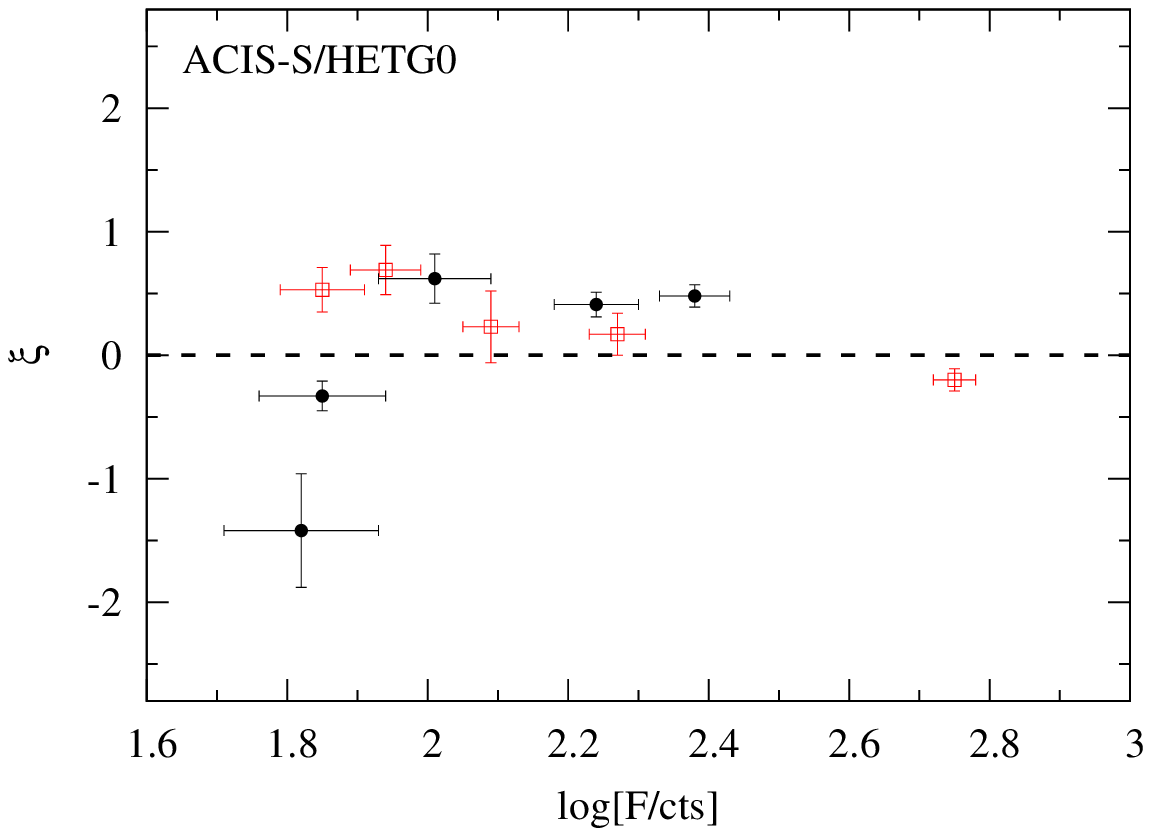}
\includegraphics[width=\columnwidth]{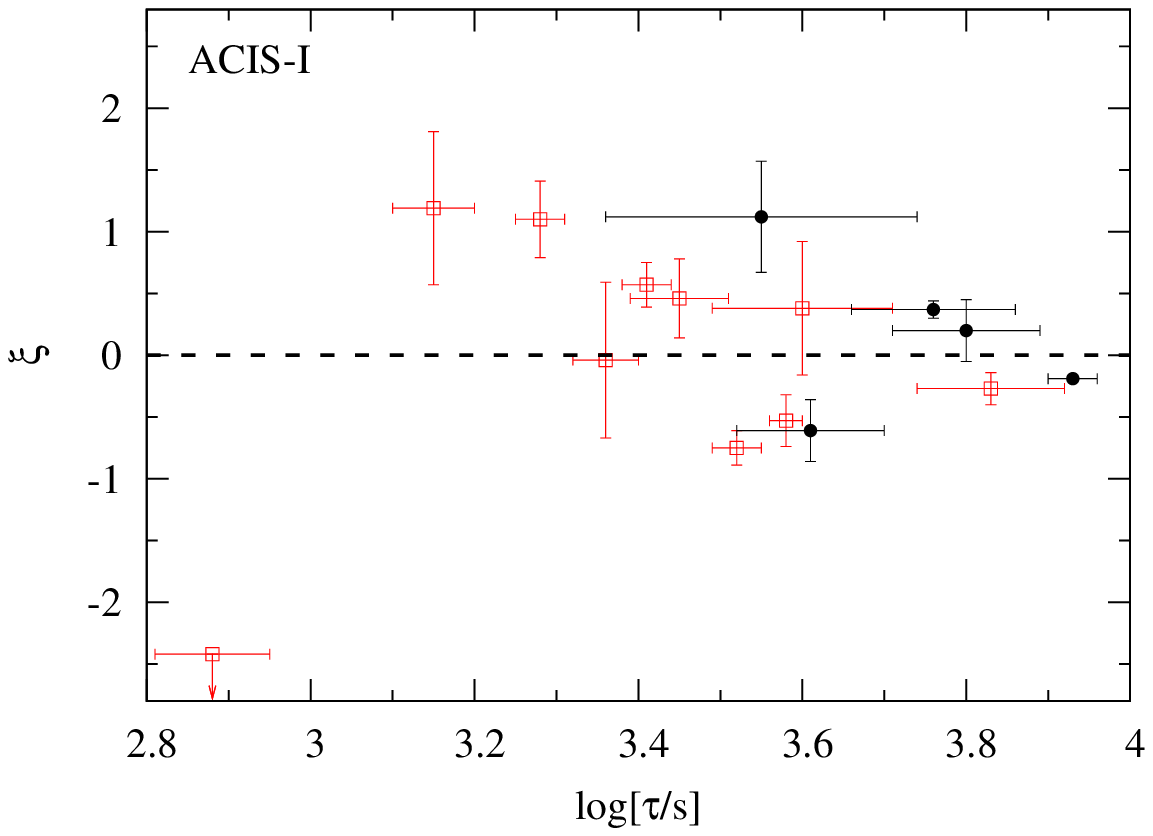}
\includegraphics[width=\columnwidth]{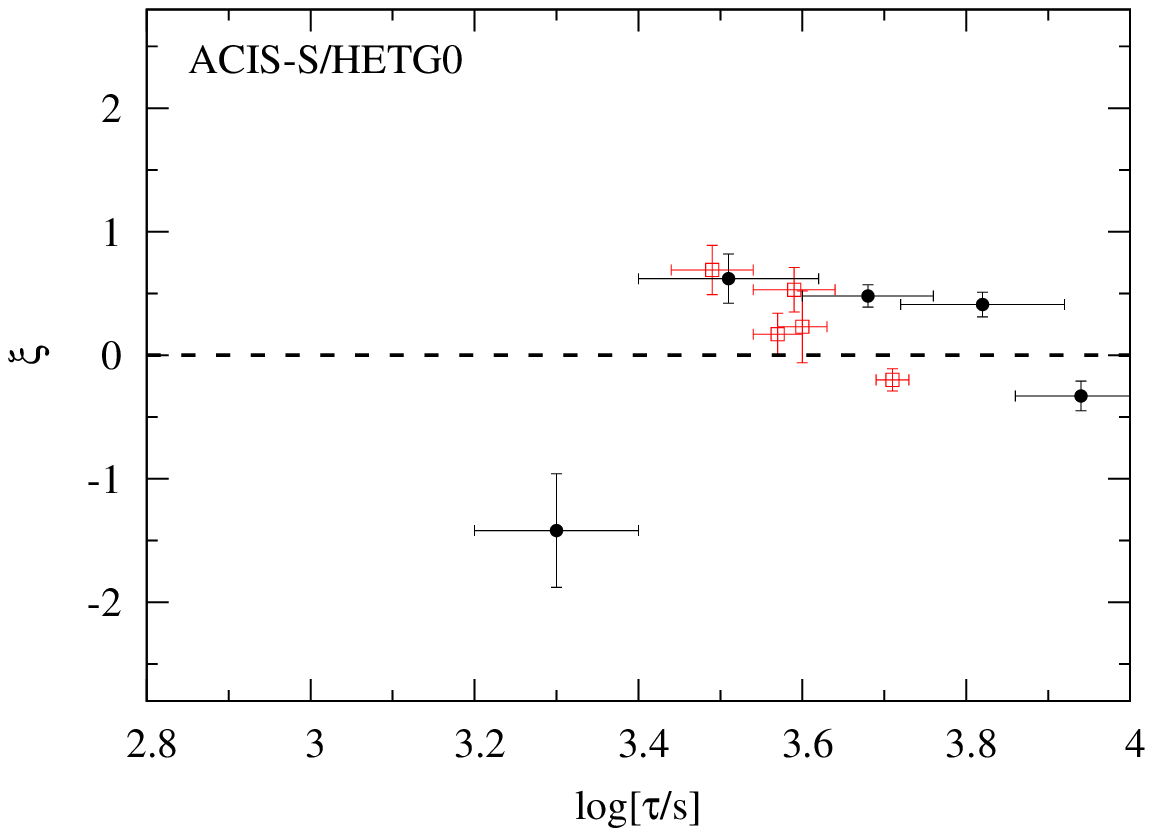}
\caption{Top panels: the profile asymmetry parameter $\xi$ versus the 
fluence for our detected flares with $F>50$ cts in the ACIS-I (left) and 
-S/HETG0 (right) samples. Bottom panels: $\xi$ versus the duration of 
the same flares. Red squares are for isolate single flares, while black 
dots are for those with apparent multiple subflare signature (see Paper I).
}
\label{fig:xi}
\end{figure*}

We refit the lightcurves of the strong flares, using the function to derive 
the shape asymmetry parameter $\xi$. For consistency, the single function 
is applied in all fits, including those with indications for subflares, 
because their effects are generally too subtle to be effectively distinguished 
from those arising from the overall profile asymmetry. The results are shown 
in Figure \ref{fig:xi},
suggesting that about half of the flares have positive $\xi$ values (hence 
fast rise and slow decay) and the other half show negative $\xi$ (slow rise 
and fast decay). The number of the flares with positive $\xi$ is only 
slightly larger that that with negative $\xi$. There is no obvious 
trend of $\xi$ with respect to the fluence. A general anti-correlation is 
present between $\xi$ and the flare durations for both samples, although 
each has one exception, which has the shortest duration among the flares.

\subsection{Flare spectral propertiers}

\begin{figure*}
\centering
\includegraphics[width=\columnwidth]{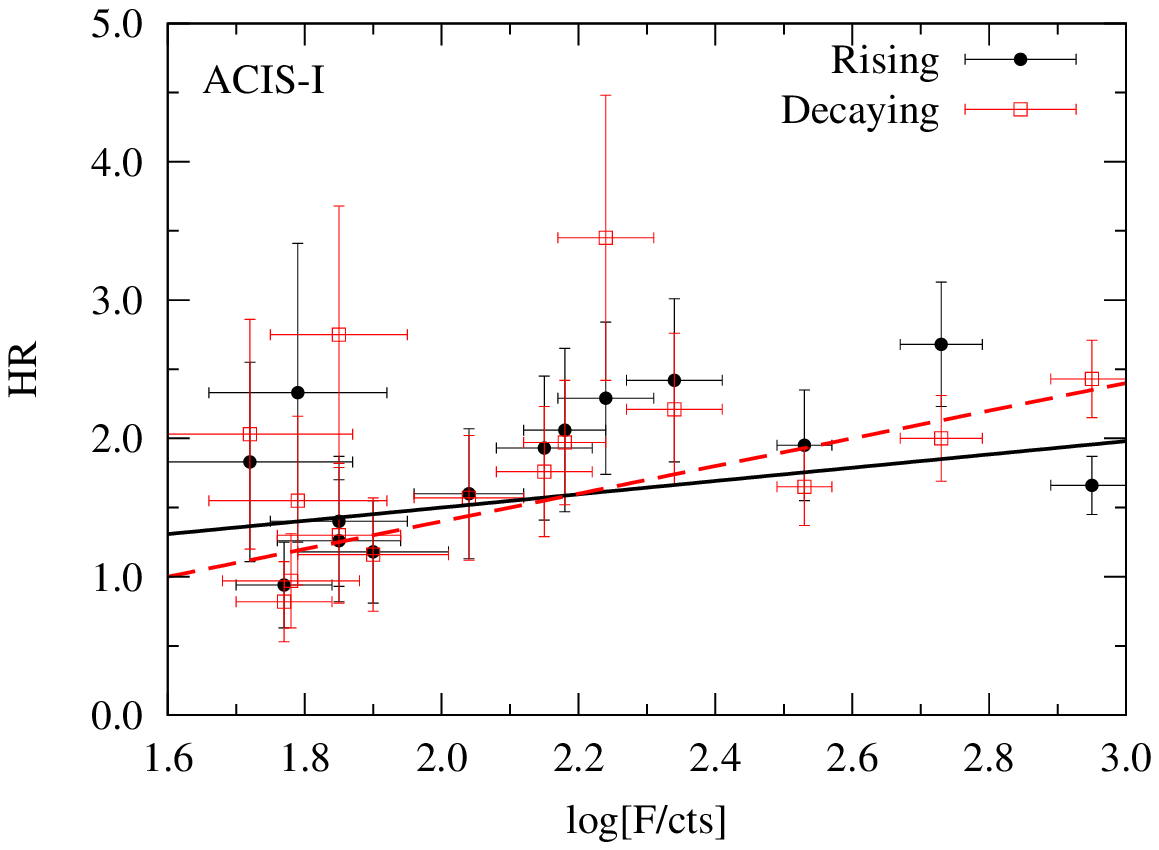}
\includegraphics[width=\columnwidth]{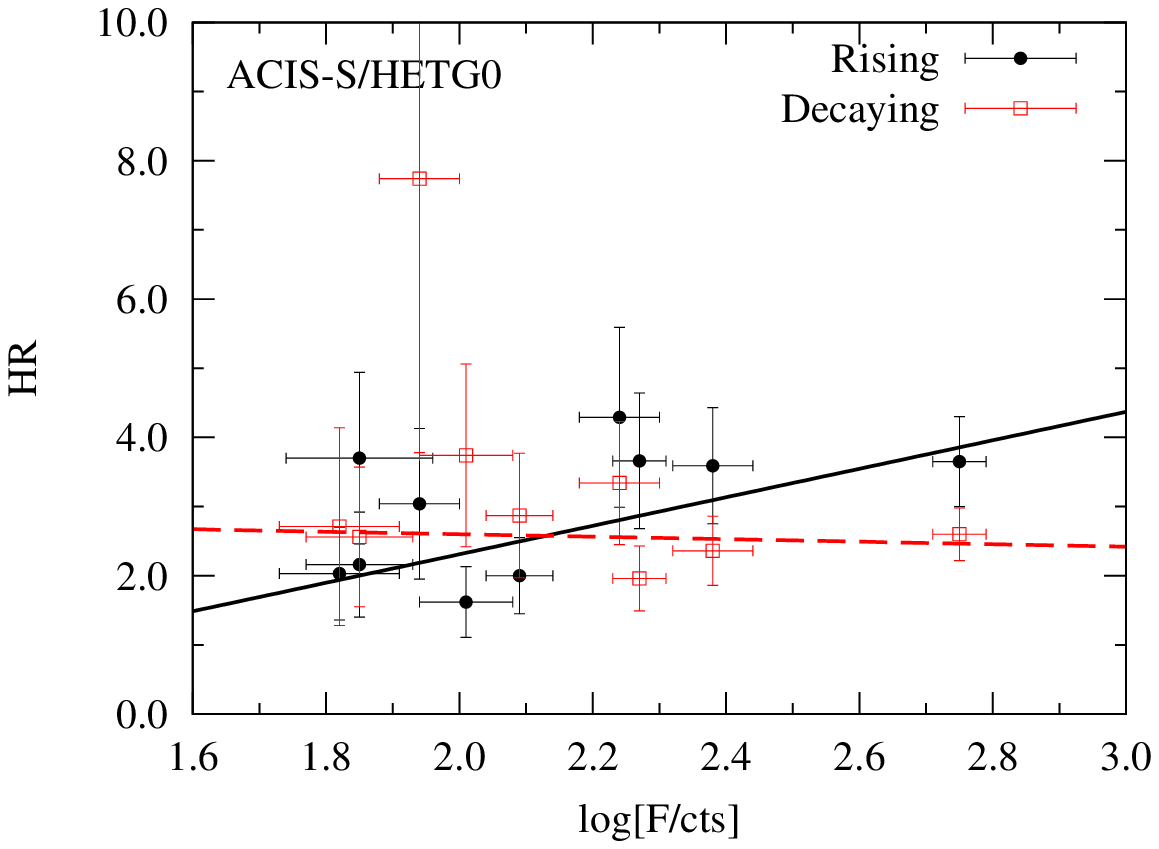}
\includegraphics[width=\columnwidth]{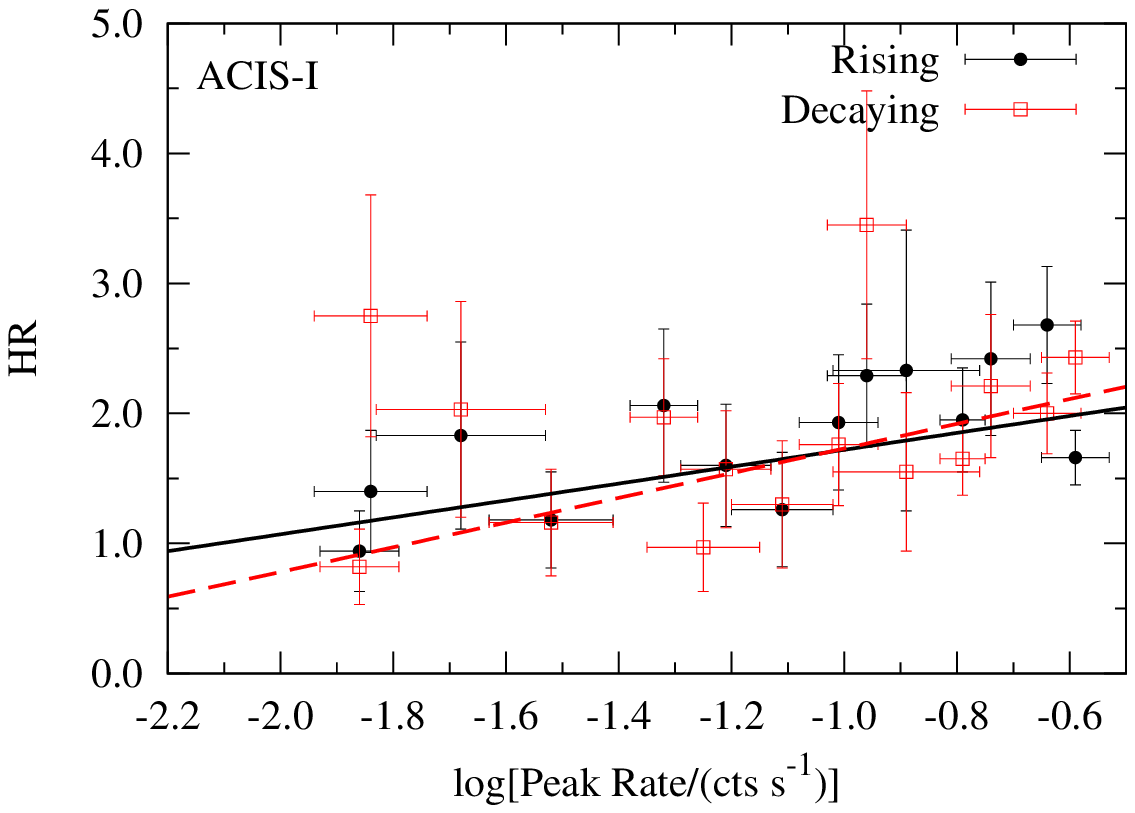}
\includegraphics[width=\columnwidth]{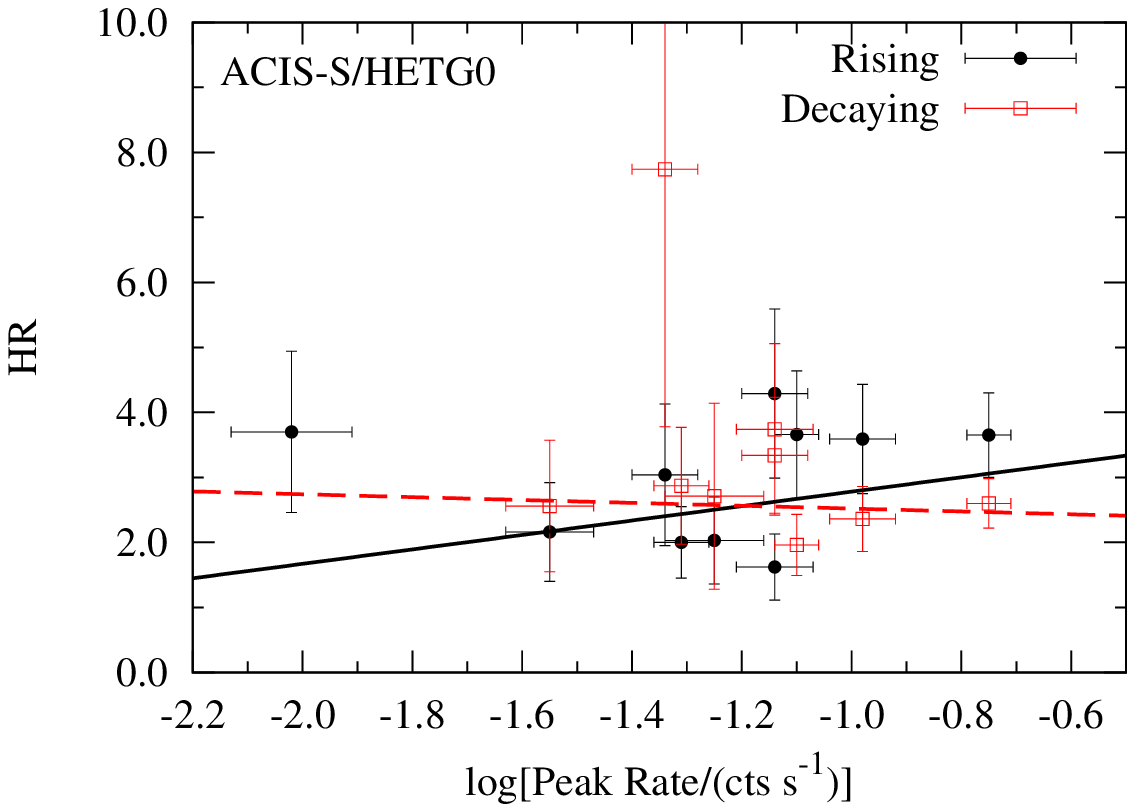}
\caption{The HR versus the fluence (top panels) or peak rate (bottom panels) 
for the flares with $F>50$ cts in the ACIS-I (left) and -S/HETG0 (right)  
samples. Lines show the linear fits characterization of the correlation 
between the two parameters for the rising and decaying phases, separately.
}
\label{fig:hr}
\end{figure*}

To characterize the spectral properties of a flare, we first define its 
spectral hardness ratio (HR) as 
\begin{equation}
{\rm HR}=\frac{N_c(4-8\,{\rm keV})}{N_c(2-4\,{\rm keV})},
\end{equation}
where $N_c$ is the number of net (quiescent contribution-subtracted) 
counts accumulated within $\pm3\sigma$ range of the Gaussian lightcurve. 
The event rate of the quiescent contribution below (above) 4 keV is 
calculated using the events detected over non-flaring time windows, 
which is 2.33 (2.55) cts/ks for the ACIS-I data and 0.73 (1.14) cts/ks 
for the -S/HETG0 data, respectively. Furthermore, to characterize the 
spectral evolution of a flare, we separate the counts into two parts, 
the rising phase before the best-fit Gaussian peak and the decaying 
phase after the peak. The results are given in Figure \ref{fig:hr}. 

We adopt a linear function, ${\rm HR}=\lambda\cdot\log F + \eta$ 
(${\rm HR}=\mu\cdot\log P + \nu$), to characterize the correlation between 
the HR and logarithmic fluence $F$ (peak rate $P$) for the two flare 
samples. The fitting results are given in Tables \ref{table:fit-hr} and
\ref{table:fit-hr2}. For the ACIS-I flares, a positive correlation 
is seen for both the rising (at a confidence level of $2\sim3\sigma$) 
and decaying phases ($\sim4\sigma$). For the ACIS-S/HETG0 flares, 
however, this correlation is less significant. Only for the rising phase 
we find a marginal correlation with a significnace of $\sim2.4\sigma$ 
($1.2\sigma$) for the HR-fluence (HR-peak-rate) correlation.
The ACIS-I data suggest that brighter flares tend to have harder spectra 
than weaker ones, especially for the decaying phase. This trend is, 
however, not obvious for the ACIS-S/HETG0 flares.

\begin{table}
\scriptsize
\centering
\caption{The best-fit values and $68\%$ uncertainties of the parameters 
characterizing the HR-fluence correlation ${\rm HR}=\lambda\cdot\log F + \eta$.}
\begin{tabular}{cccccc}
\hline \hline
    & \multicolumn{2}{c}{Rising} & & \multicolumn{2}{c}{Decaying} \\
\cline{2-3} \cline{5-6} \\
    & $\lambda$ & $\eta$ & & $\lambda$ & $\eta$ \\
\hline
ACIS-I       & $0.48\pm0.24$ & $0.54\pm0.56$ & & $1.00\pm0.24$ & $-0.60\pm0.55$ \\
ACIS-S/HETG0 & $2.06\pm0.83$ & $-1.81\pm1.77$ & & $-0.18\pm0.79$ & $2.96\pm1.90$ \\
\hline
\hline
\end{tabular}
\label{table:fit-hr}
\end{table}

\begin{table}
\scriptsize
\centering
\caption{The best-fit values and $68\%$ uncertainties of the parameters
characterizing the HR-peak-rate correlation ${\rm HR}=\mu\cdot\log P + \mu$.}
\begin{tabular}{cccccc}
\hline \hline
    & \multicolumn{2}{c}{Rising} & & \multicolumn{2}{c}{Decaying} \\
\cline{2-3} \cline{5-6} \\
    & $\mu$ & $\nu$ & & $\mu$ & $\nu$ \\
\hline
ACIS-I       & $0.65\pm0.24$ & $2.37\pm0.29$ & & $0.95\pm0.25$ & $2.68\pm0.29$ \\
ACIS-S/HETG0 & $1.11\pm0.94$ & $3.89\pm1.16$ & & $-0.22\pm1.03$ & $2.30\pm1.05$ \\
\hline
\hline
\end{tabular}
\label{table:fit-hr2}
\end{table}

We next focus on the mean spectral properties of relative faint flares, 
based on the analysis of their accumulated spectra. We limit our 
spectral analysis to those flares with negligible pile-up effects, which 
are estimated from the analysis of the lightcurves of individual flares 
in a forward fitting procedure (Paper I). In principle, correction may 
also be made in spectral fits, using the pile-up model 
\citep{2001ApJ...562..575D}, as implemented in XSPEC. However, it is not 
clear how effective the correction may be for flares, which vary strongly. 
In any case, the correction, including at least one more fitting 
parameter, would introduce additional uncertainties in the spectral 
parameter estimation \citep{2012ApJ...759...95N}. Therefore, we select 
those flares with the pile-up correction factor greater than 0.9 (i.e., 
the pile-up effect is $\lesssim 10\%$). 

We use an aperture radius of $1''.5$ to extract spectral data of \sgras. 
This extraction is made separately from the ACIS-I and -S/HETG0 
observations. We extract on-flare spectral data from the time interval 
between the $\pm3\sigma$ around the peak of each flare. If it contains 
subflares, then the interval is between their first $-3\sigma$ and last 
$+3\sigma$. We add the spectral data of individual flares together to 
form an accumulated spectrum. To examine potential flux dependent 
properties, we form two separate ACIS-S spectra from 7 strong and 37 weak 
flares, according to their individual fluences, greater or less than 
$10^{1.8} $~counts (Table~\ref{table:flare-S}). The corresponding ACIS-I 
fluence criterion is $10^{2.2}$~counts, due to the larger effective area. 
We find that all our 24 selected ACIS-I flares have fluences below this 
criterion (Table~\ref{table:flare-I}) and all have pile-up correction 
factors $< 0.9$. We further construct two off-flare spectra of \sgras, 
using the ACIS-I and -S/HETG0 data after excluding the time intervals 
of all the detected flares. These ``quiescent'' spectra are 
exposure-scaled and subtracted from the corresponding on-flare spectra 
in their analysis. 

We fit the spectra with an absorbed power-law. Specifically, th XSPEC 
model {\sl tbabs} is used to model the foreground absorption, which 
includes the contribution from dust grain \citep{2000ApJ...542..914W}, 
while {\sl xscat} to account for the grain scattering 
\citep{2016ApJ...818..143S}. The fitting is very insensitive to the 
location of the dust scattering. This parameter is thus fixed to 0.95 
(i.e., close to \sgras). A test inclusion of the {\sl pileup} model 
shows that it has little effect on the best-fitting results, 
confirming our expectation. 

\begin{figure*}
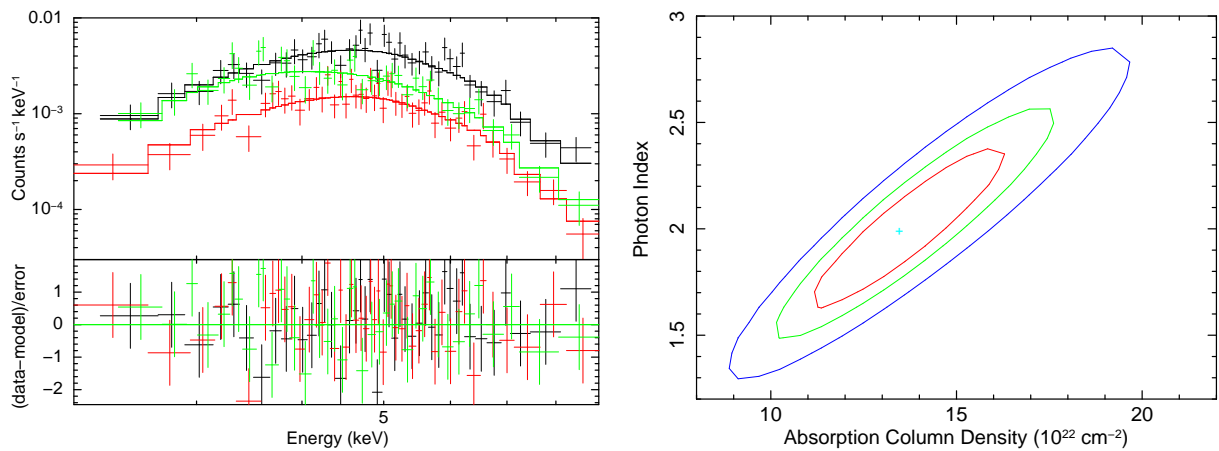

\includegraphics[width=0.7\columnwidth,angle=270]{flare_spec.ps}
\includegraphics[width=0.7\columnwidth,angle=270]{flare_spec_cont.ps}
\caption{Left panel: joint power-law model fit to the \sgras\ flare spectra. 
They are accumulated for the strong (black) and weak (red) flares detected 
with the XVP ACIS-S/HETG0 data separately, as well as the flares detected 
with the ACIS-I data (green). The spectral contributions from the 
quiescent emission, estimated from the respective data, have been subtracted. 
Right panel: 68\%, 90\%, and 99\% confidence contours of the power-law 
photon index versus the absorption column density of the fit.}
\label{f:spec_flare}
\end{figure*}

\begin{figure}
\includegraphics[width=0.7\columnwidth,angle=270]{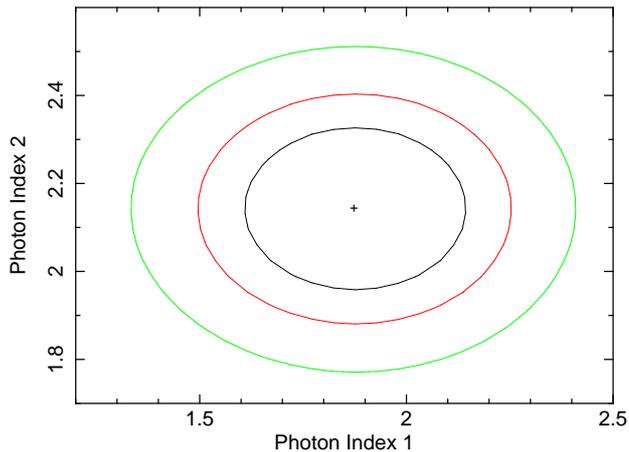}
\caption{68\%, 90\%, and 99\% confidence contours of the two photon 
indiices for the strong flares (1) and the weak ones (2).}
\label{f:spec_2index}
\end{figure}

The left panel of Figure \ref{f:spec_flare} shows that the three spectra 
of the \sgras\ flares, i.e., the weak ACIS-I flares and the strong 
and weak ACIS-S ones, can be well fitted by a single absorbed power law 
($\chi^2/n.d.f.=104/132$). The best fit photon index is $2.0\pm0.4$, and 
the absorption column density is $N_H = 13.5^{+3.1}_{-2.7}\times 10^{22}$ 
cm$^{-2}$. The uncertainties in these two parameters are largely due to 
their correlation, as shown in the right panel of Figure \ref{f:spec_flare}. 
To test any potential spectral dependence on the fluence of a flare. 
we first fix the column density to its best-fit value (i.e., removing the 
above mentioned uncertainties) and then fit the photon index for the 
strong flare spectrum independently, while keeping the indices of the 
other two spectra jointly fitted. This fit does show a marginal evidence 
that the weak flares have a slightly larger average index than that of 
the strong ones (Figure \ref{f:spec_2index}), which is consistent with 
the above HR analysis.

\section{Discussion}

The above results provide new insights on understanding the nature of the 
X-ray flare emission of \sgras\ and their origins, as well as indications 
for the possible relativistic and gravitational effects on the temporal 
and spectral properties of the flaring emission when propagating in the 
vicinity of the SMBH. We discuss these topics in the following.

\subsection{Emission mechanism}

We begin by a comparison of our spectral results with those obtained 
in previous studies, which are primarily focused on individual very bright 
flares. \citet{2017MNRAS.468.2447P} showed that the average spectral index 
of three such flares observed by \xmm\ is $\Gamma=2.20\pm0.15$. 
Similar result was found for a sample of ten flares in a wider energy
band of $1-79$ keV by \nustar\ \citep{2017arXiv170508002Z}. These results
are slightly steeper than, but still consistent within the 68\% errors 
with that obtained here. There is an indication that strong flares tend 
to have harder spectra \citep{2014ApJ...786...46B,2017arXiv170508002Z}. 
See, however, \citet{2013ApJ...769..155D} for an opposite example. 
The result obtained in this work slightly favors the former one.

Starting from a generic point of view, we may consider that the X-rays  
from a flare are predominantly generated via a single radiative process. 
Collocated particles, presumably electrons, emit the polarised NIR/IR 
synchrotron radiation. As for the X-rays, bremsstrahlung 
\citep{2002ApJ...566L..77L}, inverse Compton scattering 
\citep{2003ApJ...598..301Y,2004A&A...427....1E,2006ApJ...648.1020L, 
2008ApJ...682..373M,2012AJ....144....1Y} and synchrotron processes 
\citep{2003ApJ...598..301Y,2004ApJ...606..894Y,2009ApJ...698..676D,
2017MNRAS.468.2447P}, have been suggested as processes that give rise to 
the temporal and spectral behaviours observed in \sgras.

The bremsstrahlung requires a large emission measure, and hence a high 
plasma density in the emission region. Although it is possible for a 
local pocket of high-density plasma \citep[cf. plasmoids as in][]
{2009MNRAS.395.2183Y} to develop in an accretion inflow or outflow near 
the black hole through, for example, radiatively induced instabilities 
\citep[see][]{2002ApJ...566L..77L}, certain fine tuning is required in 
such bremsstrahlung models in order to explain the X-ray flares.   
The X-rays can also be produced when low-energy photons in the ambient 
field are Compton up-scattered by the energetic electrons that emit the 
polarised NIR/IR flare emission. During the flaring events the NIR/IR
synchrotron photons dominate the radiation field in vicinity of \sgras,
thus the X-rays are a consequence of self-Comptonisation of the synchrotron 
radiation, i.e. an SSC process. As the X-rays and the NIR radiation are 
assumed to originate from the same region, combining the data obtained in 
the NIR and X-ray observations, one can constrain the effective source size 
and the particle density \citep{2006ApJ...636..798L,2009ApJ...698..676D}. 
Analysis of a simultaneous NIR to X-ray flare by \cite{2009ApJ...698..676D}  
showed that the SSC model yielded very extreme conditions for the 
emission region: an extremely small linear size (of $\sim0.001-0.1$ 
Schwarzschild radius), a very strong magnetic field (of $\sim 10^2-10^4$ G) 
and a very high particle density (of $\sim 10^8-10^{12}\;\!{\rm cm}^{-3}$). 
The SSC model is therefore unlikely if NIR and X-ray flares are generated 
in the same location. 
  
Simultaneous observations of a very bright flare from NIR to X-ray revealed 
a spectral break between the NIR and X-ray spectra with a difference of the 
slopes $\Delta\Gamma=0.57\pm0.09$ \citep{2017MNRAS.468.2447P}. One may argue 
that this points to synchrotron radiation in the presence of radiative 
cooling. However, the result must be interpreted with caution. If the NIR 
synchrotron flares are produced by the same population of electrons that 
are injected into the emission region as the X-ray ones and no efficient
particle escape, we would expect a delay of NIR emission with respect to 
the X-ray one on the radiative cooling timescale. The observations do not 
support such a delay \citep{2017MNRAS.468.2447P}. 

For a homogeneous emission region with a single instantaneous particle 
injection, the effective cooling time can be estimated from the observed 
peak of the radiative spectrum $\nu_{\rm m}$, as $\tau_{\rm cool} = 
5\times10^{11}(B\ \langle \sin \alpha \rangle)^{-3/2}\nu_{\rm m}^{-1/2}
~{\rm sec}$ \citep[see][]{1975rpa..book.....T}, where $B$ is the magnetic 
field threading the region and $\alpha$ is the pitch angle of the electrons 
with respect to the magnetic field. If we assume that $B\sim 10~{\rm G}$ 
\citep{2009ApJ...698..676D} and the electron momentum distribution is 
isotropic, for $\nu_{\rm m}\sim10^{18}~{\rm Hz}$ we have $\tau_{\rm cool}
\sim 0.75$ min. As the cooling time is much shorter than the duration of a 
flare, the acceleration (or injection) of electrons therefore cannot be due 
to an impulsive single event. The flare's variability is therefore caused 
by the dynamical evolution of the system, with temporal variations in the 
injection process, if a single emission region dominates. Alternatively, 
spatial propagation of magnetic eruption fronts will lead to multiple 
injection/acceleration sites, giving rise to multiple emission regions. 

Our analyses show no significant difference in the HRs between the rising 
and decaying phases (Figure~\ref{fig:hr}), which does not support the 
shutdown of the flare being due to synchrotron cooling in a uniform plasma,
because of the short cooling timescale and the anticipated dramatic spectral 
softening. Such persistence of the HR is however allowed, if the radiative 
particles escape from the region or the magnetic field dissipates. It is 
also allowed if the system is dynamical, with multiple particle 
injection/acceleration episodes and/or continuous particle 
injection/acceleration along a propagating magnetic reconnection front.   

\subsection{Origin of flares}

We compare our improved statistical constraints on the fluence and duration
distributions of the X-ray flares with the predictions of the various 
scenarios for the generation of \sgras\ X-ray flares. Among the broad class 
of magnetic reconnection scenarios for eruptive flares, SOC is a variant 
of the phenomenological models allowing a propagating front. The flare 
statistics in an SOC model depends on the effective geometric dimension 
of the system. For instance, a classical diffusion model predicts 
$\alpha_{\rm E}=3/2$ for the total energy (or the fluence) distribution, 
$\alpha_{\rm T}=2$ for the duration distribution, and $\beta=1/2$ for the 
duration-fluence correlation, for the spatial dimension of $S=3$ 
\citep{2016SSRv..198...47A}. The observations of solar flares give on 
average $\alpha_{\rm E}=1.62\pm0.12$ and $\alpha_{\rm T}=1.99\pm0.35$, 
which are well consistent with the SOC predictions with $S=3$ 
\citep{2016SSRv..198...47A}. 

The (joint) statistical analysis of the X-ray flares in \S~2.1 reveals that 
the fluence distribution slope is $\alpha\sim1.7$, with the $95\%$ 
lower limit of $1.54$, which is considerably larger than the prediction of 
the simple SOC model for $S=3$. The duration versus fluence correlation 
is found to be very weak ($\beta\sim0$). The $95\%$ upper limit
of $\beta$ is about 0.23, which is substantially smaller than that (0.5) 
expected from the classical fractal diffusive SOC model. These results 
imply that the X-ray flares may not be self-similar, as predicted by 
the simple SOC model. It is possible that the non-uniform scenario of the 
SOC model with, e.g., finite boundary conditions, is responsible for such 
distributions of the flares. Alternatively, the X-ray fluence may not be
a good measurement of the total energy of a flare.

A very different scenario for the production of \sgras\ flares is the tidal
disruption of asteroids by the SMBH \citep[e.g.,][]{2008A&A...487..527C,
2009A&A...496..307K,2012MNRAS.421.1315Z}. Asteriods could be split into
small pieces when passing close enough (e.g., within 1 AU) by the SMBH.
They may then be vaporized by bodily friction with the accretion flow. 
A transient population of high-energy particles may be produced via the
shock due to the bulk kinetic energy of an asteroid and/or plasma 
instabilities, leading to a flare of radiation \citep{2012MNRAS.421.1315Z}. 
This asteroid disruption and evaporation model explains the luminosities, 
time scales and event rates of the flares, at least on the orders of 
magnitude. There is so far no clear prediction for the fluence distribution 
as well as the fluence-duration correlation of the model. However, 
in a very simple and rough analogy of the Galactic center 
environment to the Oort cloud of the solar system, one may assume that 
the size distribution of asteroids can be characterized by a power-law, 
$dn(r)/dr\propto r^{-q}$, with $q\sim 3-4$ \citep{2012MNRAS.421.1315Z}. 
The fluence distribution of the flares simply follows the mass function 
of asteroids, which is $dn/dM\propto M^{(-q-2)/3}$. Therefore we have 
$\alpha\sim 1.7-2$, which is consistent with that obtained in our 
analysis (see Table \ref{table:fit}). The typical duration of a 
flare is then determined by the flyby time of the asteroid, which is
independent of the asteroid size \citep{2012MNRAS.421.1315Z}. 
The predictions of the model are thus consistent with our observations.
More detailed modeling of the asteroid distribution in the Galactic 
center environment, as well as the disruption and radiation processes 
of this scenario, is needed to further test its viability.

\subsection{SMBH environment effect on the flare profile}

Most of astronomical flaring events, such as the soft X-ray and lower-energy 
emission from $\gamma$-ray bursts \citep[GRBs;][]{1995ARA&A..33..415F} and 
(low energy) solar flares \citep{2011SSRv..159...19F}, show ``fast rise and 
slow decay'' lightcurves (i.e., $\xi>0$), revealing the fast acceleration 
and slow depletion \citep[via e.g., cooling or escape;][]{2017MNRAS.468.2552L} 
of particles. Our analysis 
of the flare profiles of \sgras\ in \S~2.2 shows that almost half of the 
flares have such common ``fast rise and slow decay'' lightcurves and 
the other half are opposite, which is analogous to the impulsive component 
of the hard X-rays and higher energy emission of solar flares and GRBs.  
This result may also indicate that the observed lightcurves are not 
intrinsic and may result from radiation propagation in the extreme 
environment of the SMBH. The general anti-correlation between $\xi$ and 
$\log\tau$ as shown in Figure \ref{fig:xi} supports this picture. 
Intrinsically flares are most likely produced with shorter durations and 
``fast rise slow decay'' profiles. The observed broader and diverse 
lightcurves may largely result from the gravitational lensing and Doppler 
effects due to the orbital motion and/or the general relativity frame 
dragging. These effects tend to smear the lightcurve of a flare, giving 
less distinct sub-structures of its profile \citep{2015MNRAS.454.3283Y}. 
The effects also depend on the flare starting position relative to the 
black hole and increase with the inclination angle of the accretion flow 
and with the spin of the SMBH. Furthermore, the effects are energy-dependent, 
which may be used to distinguish them from the intrinsic properties of 
flares. Therefore, with sufficient counting statistics and energy coverage 
of observations, \sgras\ X-ray flares can, in principle, be used to probe 
the spin and the space-time structure around the event horizon of the 
SMBH, as well as the inclination angle of the innermost accretion disk. 

\section{Summary}

We have studied the statistical properties of a sample of 82 flares 
detected in the \chandra\ observations from 1999 to 2012 (Paper I). 
In the analysis of the flare fluences and their correlation with the 
durations, we use the MCMC technique to forward fit model parameters,
accounting for both detection incompleteness and bias, which are found 
to be very important. We further systematically analyze the lightcurve
asymmetry and spectral HR of individual bright flares with fluences $>50$ 
counts, as well as the accumulared spectra of relatively weak flares. 
We summarize our major findings as follows.
\begin{itemize}

\item The fluence distribution can be well modeled by a power-law with
a slope of $1.73^{+0.20}_{-0.19}$, which is inconsistent with the prediction 
of $1.5$ from the simple classical fractal diffusive SOC model with 
geometric dimension $S\lesssim3$.

\item There is no statistically significant correlation between the flare
fluence and duration, which is again inconsistent with the prediction of 
the simple SOC model. The intrinsic duration dispersion of the flare is 
about $0.3$ dex around the best-fit power-law relation. 

\item About half of the relatively bright flares show ``fast rise and 
slow decay'' profiles, whereas the other half are opposite. This is 
different from the commonly observed ``fast rise and slow decay'' profiles 
from astrophysical transients, such as GRBs and solar flares, indicating 
that the flare shape may not be intrinsic. The gravitational lensing 
and Doppler effects of the flare radiation around the SMBH may play a
dominant role in regulating the shape.

\item The accumulated spectra of the flares can be well characterized by a
power-law of photon index $\Gamma=2.0\pm0.4$. We find a marginal trend that 
the spectra of brighter flares are harder than those of relatively weak ones. 
No significant HR difference between the rising and decaying phases of 
the X-ray flares is found.

\end{itemize}

While these results provide new constraints on the origin of \sgras\ flares,
as well as their X-ray emission mechanism, more detailed modeling of their
production and evolution is clearly needed. In particular, dedicated 
simulations of photons traveling through the space and time, strongly 
affected by the presence of the SMBH and the resulting flare shapes will
be useful for comparison with the observations. Such comparison will
provide important tests on various scenarios for the production of the
X-ray flares and a potential tool to measure the spin of the SMBH.

\section*{Acknowledgments}

We thank the referee for constructive comments, which helped to improve 
the presentation of the paper. QY is supported by the 100 Talents program 
of Chinese Academy of Sciences. QDW acknowledges the support of NASA via 
the SAO/CXC grant G06-17024X.

\bibliographystyle{mn2e}
\bibliography{refs}

\end{document}